   \documentclass[fleqn,useAMS,usenatbib]{mnras}

    \usepackage{color,subfigure,lineno} 
    
    \newcommand{\publishMode}[0]{} 

    \ifx\publishMode\undefined
       
    \else
      
    \fi 

   \newcommand{\hbeta}{H{$\beta$}}
   \def\CIV{C\,{\sc iv}}
   \def\MgII{Mg\,{\sc ii}}

   \newcommand{\lya}{$\mathrm{Ly}\alpha$ }
   \newcommand\addtag{\refstepcounter{equation}\tag{\theequation}}
   \usepackage{hyperref}
   \usepackage{listings}
   \usepackage{chngcntr}

  	\usepackage{soul,xcolor}
  	\usepackage{ulem}
  	\usepackage{graphicx}
  	\usepackage{amsmath,amssymb}
  	\usepackage{nicefrac}
  	\usepackage{rotating}
  	\usepackage{pdflscape}	
  	\usepackage{makecell}
  	\usepackage{booktabs,caption}
  	\usepackage[flushleft]{threeparttable}
  	
  	\usepackage{dcolumn}
  	\newcolumntype{d}[1]{D{.}{.}{#1}}
	\newcommand\plotone[1]{%
	  	\centering 
	  	\includegraphics[keepaspectratio,width=0.85\columnwidth]{#1}%
	}%
	\newcommand\plottwo[2]{%
	  	\centering 
	  	\includegraphics[keepaspectratio,width=0.85\columnwidth]{#1}%
	  	\hfil 
	  	\includegraphics[keepaspectratio,width=0.85\columnwidth]{#2}%
	}%

	\usepackage[T1]{fontenc}
	\usepackage{ae,aecompl}
	
	\usepackage{newtxtext,newtxmath}

	\title[Demographics of Quasars at $z\sim 6$]{Demographics of $z\sim 6$ Quasars in the Black Hole Mass-Luminosity Plane}    
	\author[Wu et al.]{
		Jin Wu,$^{1,2}$\thanks{e-mail: \href{mailto:wujin\_astro@pku.edu.cn}{wujin\_astro@pku.edu.cn}}
		Yue Shen,$^{3,4}$
		Linhua Jiang,$^{1,2}$\thanks{e-mail: \href{mailto:jiangKIAA@pku.edu.cn}{jiangKIAA@pku.edu.cn}}
		Eduardo Ba\~{n}ados,$^{5}$
		Xiaohui Fan,$^{6}$
		Luis C. Ho,$^{1,2}$
		\newauthor
		Marianne Vestergaard,$^{7}$
		Feige Wang,$^{6}$
		Shu Wang,$^{8}$
		Xue-Bing Wu,$^{1,2}$
		Jinyi Yang$^{6}$
		\\
		$^{1}$Kavli Institute for Astronomy and Astrophysics, Peking University, Beijing 100871, China\\
		$^{2}$Department of Astronomy, School of Physics, Peking University, Beijing 100871, China\\
		$^{3}$Department of Astronomy, University of Illinois at Urbana-Champaign, Urbana, IL 61801, USA\\
		$^{4}$National Center for Supercomputing Applications, University of Illinois at Urbana-Champaign, Urbana, IL 61801, USA\\
		$^{5}$Max-Planck-Institut f\"{u}r Astronomie, K\"{o}nigstuhl 17, D-69117, Heidelberg, Germany\\
		$^{6}$Steward Observatory, University of Arizona, 933 North Cherry Avenue, Tucson, AZ 85721, USA\\
		$^{7}$The Niels Bohr Institute at University of Copenhagen, Juliane Maries Vej 30, DK-2100 Copenhagen, Denmark\\
		$^{8}$Astronomy Program, Department of Physics and Astronomy, Seoul National University, Seoul, 08826, Republic of Korea
	}

\begin{document}
    \pagerange{\pageref{firstpage}--\pageref{lastpage}}
    \maketitle


    \begin{abstract} 

    We study the demographics of $z\sim 6$ broad-line quasars in the black hole (BH) mass-luminosity plane using a sample of more than 100 quasars at $5.7<z<6.5$. These quasars have well quantified selection functions and nearly one third of them also have virial BH masses estimated from near-IR spectroscopy. We use forward modeling of parameterized intrinsic distributions of BH masses and Eddington ratios, and account for the sample flux limits and measurement uncertainties of the BH masses and luminosities. We find significant differences between the intrinsic and observed distributions of the quantities due to measurement uncertainties and sample flux limits. There is also marginal evidence that the virial BH masses are susceptible to a positive luminosity-dependent bias (BH mass is overestimated when luminosity is above the average), and that the mean Eddington ratio increases with BH mass. Our models provide reliable constraints on the $z\sim 6$ black hole mass function at $M_{\rm BH}>10^{8.5}\,M_\odot$, with a median $1\sigma$ uncertainty of $\sim$ 0.5 dex in abundance. The intrinsic Eddington ratio distribution of $M_{\rm BH}>10^{8.5}\,M_\odot$ quasars can be approximated by a mass-dependent Schechter model, with a broad peak around $\log(L_{\rm bol}/L_{\rm Edd})\sim -0.9$. We also find that, at $4.5 \lesssim z \lesssim 6$, the number densities of more massive BHs tend to decline more rapidly with increasing redshift, contrary to the trend at $2.5 \lesssim z \lesssim 4.5$ reported previously. 
    \end{abstract}
    
    \begin{keywords}
    quasars: supermassive black holes -- methods: statistical
  	\end{keywords}

\section{Introduction}
\label{sec:introduction}

The cosmic evolution of supermassive black holes (SMBHs) is among the most important topics of galaxy formation and evolution. It is now widely accepted that SMBHs reside in the centers of massive galaxies, and they grow by gas accretion during which they are witnessed as active galactic nuclei (AGNs) or quasars, the most luminous subset of AGNs. The past few decades have seen significant progress in observations of quasars across cosmic time and in theoretical studies of the growth and evolution of SMBHs \citep[e.g.,][]{Haiman_Loeb_1998, Gebhardt_etal_2000, Kauffmann_Haehnelt_2000, Volonteri_etal_2003, Hopkins_etal_2008, Somerville_etal_2008, Gultekin_etal_2009, Shankar_etal_2009}. Quasars have now been discovered up to $z>7$, with SMBH masses $\gtrsim 10^9\,M_\odot$ \citep[e.g.,][]{Mortlock_etal_2011, Wu15QSO, Jiang2016SDSSQ, Banados_etal_2018, Yang2020, Wang_etal_2021}. While the physical properties of these $z>6$ quasars are similar to their lower-redshift counterparts \citep[e.g.,][]{Shen_etal_2019b, Yang2021}, it is unclear how these SMBHs could have grown to such large masses when the Universe was less than one billion years old \citep[e.g.,][]{Volonteri_2010,Haiman_2013,Inayoshi_etal_2020}. 

In the meantime, massive wide-field spectroscopic surveys have compiled large quasar samples at $0<z\lesssim 5$, enabling robust measurements of the quasar abundance across most of the cosmic history \citep[e.g.,][]{Schneider_etal_2010, Richards2006SED, Lyke_etal_2020}. The demographics of quasars contains crucial information about the cosmic assembly of SMBHs. The basic demography is the quasar luminosity function (QLF), i.e., the space density of quasars at different luminosities. With ever increasing sample statistics from massive sky surveys, QLF has been measured for different AGN populations in different bands
\citep[e.g.,][]{Boyle_etal_2000, Fan_etal_2001, Croom_etal_2004, Hao_etal_2005, Richards2006SED, Silverman_etal_2008, Croom_etal_2009, Willott2010QLF, McGreer_etal_2013,Ross_etal_2013, Yang2016, McGreer2018, WangFeige2019b, Kim2020QLF}.
With spectroscopy, one can also estimate the black hole (BH) masses of these quasars using the so-called ``single-epoch virial BH mass estimators'' \citep[e.g.,][]{Vestergaard_Peterson_2006,Shen2013BHM}, enabling the measurements of the BH mass function (BHMF) in quasars \citep[e.g.,][]{GreeneHo2007BHMF, Vestergaard_Osmer_2009,Schulze2010BHMF,ShenKelly2012BHMF,KellyShen2013BHMF}. Either QLF or BHMF is the 1D projection of the 2D quasar distribution in the mass-luminosity plane with condensed information. An alternative approach is to constrain the abundance of quasars in the mass-luminosity plane directly, which provides more information about the accretion properties of these SMBHs (e.g., the Eddington ratio) and better constraints on their cosmic evolution \citep[e.g.,][]{ShenKelly2012BHMF,KellyShen2013BHMF}. 

In this work, we constrain the abundance of $z\sim 6$ quasars in the mass-luminosity plane using the largest homogeneous sample available at $z\sim 6$ \citep{Willott2010QLF, Jiang2016SDSSQ, Matsuoka2018QLF}. We conduct a forward modeling approach that carefully accounts for the selection functions and uncertainties in the measured quantities, particularly the uncertainties and biases in BH mass estimates. This paper is organized as follows. In \S\ref{sec:sample}, we describe our quasar samples constructed from the SDSS \citep{Jiang2016SDSSQ}, CFHQS \citep{Willott2010QLF}, and SHELLQs \citep{Matsuoka2018QLF} high-$z$ quasar surveys. Two nearly complete BH mass samples are also constructed from the SDSS survey. In \S\ref{sec:method}, we describe our methodologies of measuring the 1D BHMF and Eddington ratio distribution function (ERDF), as well as the 2D distribution in the mass-luminosity plane, using three methods: the $1/V_{\rm max}$ method, the maximum likelihood method, and the QLF fitting method. In \S\ref{sec:result}, we present our results. We discuss the implications of our results in \S\ref{sec:discussion} and summarize the paper in \S\ref{sec:summary}.

Throughout this paper, we adopt a flat $\Lambda$CDM cosmology with cosmological parameters of $( \Omega_M , \Omega_\Lambda , H_0) = (0.3, 0.7, 70~\mathrm{km}~\mathrm{s}^{-1}~\mathrm{Mpc}^{-1})$. We use the symbol $M_{\mathrm{BH}}$ (or $M_{\bullet}$), $\lambda$ (or $\lambda_{\mathrm{Edd}}$), and $L_{\mathrm{bol}}$ to denote the BH mass, Eddington ratio, and bolometric luminosity, respectively. For simplicity in the presentation of formalism, we use the same symbols, e.g., $\Phi(...)$ or $\Psi(...)$, to denote probability distribution functions of different quantities, explicitly specified in the parentheses, e.g., $\Phi(M_\bullet)$ denotes the 1D BHMF. Distribution functions by default are defined per logarithmic intervals; however, when needed we convert the distribution function to that on linear intervals without introducing new symbols -- this technical detail is properly implemented in our calculations.

We distinguish three types of probability distributions of abundance in our demographics modeling. The ``intrinsic'' distribution refers to the true, underlying distribution without being modified by sample selection functions and measurement uncertainties in physical quantities. The ``measured'' distribution refers to the intrinsic distribution convolved with the error model to account for measurement uncertainties, but not affected by the selection functions. Finally, the ``observed'' distribution refers to the final observed distribution impacted by both selection functions and measurement uncertainties.

\section{Quasar Samples}
\label{sec:sample}

We consider two different categories of samples in this study. The first category includes quasars with luminosity measurements only, and are referred to as luminosity samples. The other category of samples include quasars with both luminosity and BH mass measurements, and are referred to as BH mass samples. There are generally more quasars in luminosity samples than in BH mass samples. Both categories of samples have well defined selection functions that are required to perform statistical analysis. For either category, there are multiple samples collected from the literature, as detailed below. The BH mass samples are used to jointly constrain the 2D distribution in the mass-luminosity plane, while the larger luminosity samples are used to improve the constraints on the BHMF (\S\ref{sec:method}).

\subsection{Luminosity Samples}
\label{thelfsamples}

We start with the quasar sample compiled in  \citet{Matsuoka2018QLF}. It contains 110 quasars at $5.7 \leqslant z \leqslant 6.5$, covering the magnitude range of $-22 < M_{1450} < -30$. These quasars are drawn from the SDSS, CFHQS and SHELLQs high-$z$ quasar surveys, respectively, which roughly cover the bright, intermediate, and faint luminosity regimes of $z\sim 6$ quasars. The information of the quasar samples used in this work is summarized in \autoref{tab:all-samples}.

There are 52 quasars from the SDSS survey at $5.7 \leqslant z \leqslant 6.5$, and 47 of them were used to construct a statistically complete sample ~\citep{Jiang2016SDSSQ}. They belong to three nested high-$z$ quasar searches using the SDSS data. The first one is the main survey with a flux limit of $z_{\rm AB}\sim20$ mag that used single-epoch SDSS imaging for target selection. The second one is the overlap region survey that used two or more repeated photometric observations of overlapping fields in the SDSS \citep{Jiang2015SDSSO}; the flux limit is $z_{\rm AB}\sim20.5$ mag. The last one is the Stripe 82 survey that used deep co-added data from repeated observations in the SDSS Stripe 82 region \citep{Jiang2014S82}; the flux limit is $z_{\rm AB}\sim22$ mag.

There are 24, 17, and 13 quasars from the SDSS main, overlap, and Stripe 82 surveys with different flux limits ($\gtrsim 10\sigma$ detection in the $z$ band). After removing 7 duplicate quasars in more than one surveys, the final SDSS sample contains 47 unique quasars. These SDSS quasars span a luminosity range from $M_{1450}=-30$ to $-24$ mag and represent the most luminous quasars at $z \sim 6$. The effective areas of the main, overlap, and Stripe 82 surveys are 11,240, 4,223, and 277 $\mathrm{deg}^2$, respectively. The selection functions of the SDSS quasars are shown in Figure 6 in \citet{Jiang2015SDSSO}.

We use the complete sample of 17 quasars at $5.7 \leqslant z \leqslant 6.5$ from the CFHQS survey \citep{Willott2010QLF}. This sample contains 16 quasars with $M_{1450}=-27$ to $-24$ mag in a wide-area survey and 1 quasar with $M_{1450}=-22.2$ mag in a deep survey. The effective areas of the two surveys are 494 and 4.47 $\mathrm{deg}^2$, respectively. The selection functions of the CFHQS quasars are shown in Figure 4 in \citet{Willott2010QLF}.

We use 48 SHELLQs quasars at $5.7 \leqslant z \leqslant 6.5$ from the HSC-SSP Wide survey ~\citep{Matsuoka2018QLF}, selected by a single set of criteria with $z_{\mathrm{AB}}<24.5$ mag and $i_{\mathrm{AB}}-z_{\mathrm{AB}}>2.0$ mag. The luminosity range is $-25.5 \leqslant M_{1450} \leqslant -22.5$ and the total effective area is 646 $\mathrm{deg}^2$. This sample probes a lower-luminosity range of $z\sim 6$ quasars than the SDSS and CFHQS samples. The selection functions of the SHELLQs quasars are shown in Figure 9 in \citet{Matsuoka2018QLF}.

For quasars in the luminosity samples, their absolute magnitudes ($M_{1450}$) are estimated by extrapolating the continuum spectrum redward of \lya to rest-frame 1450{\AA}, assuming a fixed power-law continuum $f_\lambda \propto \lambda^{-1.5}$. The selection functions were based on the optical and near-IR flux and were determined by simulations in the corresponding surveys. They were calculated by applying target selection criteria to mock quasar spectra in the ($M_{1450}, z$) plane with reasonable dispersions in the mock spectra (e.g., continuum shapes; see the original survey papers for details) and computing the selection probability at each grid point.
More details about these luminosity samples are presented in  \citet{Matsuoka2018QLF}.
The number of quasars in each sample is summarized in \autoref{tab:all-samples}.

\subsection{BH Mass Samples}
\label{sec:thebhmasssamples}

Quasar BH masses can be estimated from single-epoch spectra using virial mass estimators ~\citep[e.g.,][]{Vestergaard_Peterson_2006,Shen2013BHM}. This method employs the virial relation $M_\bullet=V^2_{\mathrm{vir}} R/G=f~W^2 R/G$, where $f$ is the scaling factor that accounts for the structure and the unknown inclination angle of the AGN broad-line region, $V_{\mathrm{vir}}$ is the virial velocity of the broad line clouds, $W$ is the broad emission line width \citep[e.g.,][]{Wang_etal_2019}, and $R$ is the size of the broad-line region. From local AGN reverberation mapping observations \citep[e.g.,][]{Du2016, Fonseca_Alvarez_etal_2020}, there is a correlation between the measured $R$ and the monochromatic optical continuum luminosity $L$ ~\citep[the $R-L$ relation; e.g.,][]{Peterson2010RM, Bentz2013}. Assuming that this scaling relationship is applicable to broad-line quasars at high redshifts and high luminosities, we can estimate BH masses using broad line widths and continuum luminosities measured from single-epoch spectra. We refer the reader to \citet{Shen2013BHM} for detailed discussions on the uncertainties and caveats in these single-epoch virial masses for high-redshift quasars.

To obtain virial BH masses for $z\sim 6$ quasars, near-IR spectroscopy is necessary to cover rest-frame UV broad lines such as \MgII\ and \CIV.  \citet{Shen2019GNIRS} conducted a large Gemini program to acquire GNIRS near-IR spectroscopy for $z \gtrsim 5.7$ quasars, and presented virial mass estimates for 50 objects, in which 29 objects are SDSS quasars. Here we adopt virial BH masses estimated from \MgII\ if available, otherwise \CIV-based masses are used. 

The virial BH mass estimate can be expressed as
\[
\log({\frac{M_{\mathrm{BH,vir}}}{M_\odot}}) = a+b\log({\frac{\lambda L_\lambda}{10^{44}\mathrm{erg}\ \mathrm{s}^{-1}}})+2\log({\frac{\mathrm{FWHM}}{\mathrm{km}\ \mathrm{s}^{-1}}})
\]
where the coefficients $a$ and $b$ are empirically calibrated against local measurements. We use $(a,b) = (0.740, 0.62)$ \citep{Shen2011C} for \MgII\ and $(a,b) = (0.660, 0.53)$ \citep{Vestergaard_Peterson_2006} for \CIV\ as our fiducial BH mass recipes, which provide consistent mass estimates for high-luminosity quasars if both lines are available \citep{Shen2011C}. 

We calculate bolometric luminosities using a constant bolometric correction of 5.15 ~\citep{Richards2006SED} from the 3000\,\AA\ monochromatic luminosities. The Eddington ratio is calculated as $\lambda_{\mathrm{Edd}}=L_{\mathrm{bol}}/L_{\mathrm{Edd}}$, where $L_{\mathrm{Edd}}=1.3\times10^{38} \mathrm{erg s}^{-1}(M_{\mathrm{BH}}/M_{\odot})$ is the Eddington luminosity of the black hole. Recently, \citet{WangShu_etal_2021} reprocessed this GNIRS sample in \citet{Shen2019GNIRS} to study the broad-line region metallicity of $z\sim 6$ quasars.
They updated measurements for 23 SDSS quasars in the \citet{Shen2019GNIRS} sample, which we adopt here as our fiducial BH masses (see \autoref{tab:bhm-sample} for details). These masses are consistent with those reported in \citet{Shen2019GNIRS} within 1$\sigma$. This GNIRS sample is the main BH mass sample for our demographic study. We also collected the near-IR spectral fitting results of 9 SDSS quasars from the literature~\citep{Jiang07BHM, DeRosa2011BHM, Wu15QSO, Schindler2020} and recalculated their virial BH masses using the same BH mass recipes. In total, there are 38 SDSS quasars with BH mass measurements.

For reliable modeling of the sample, we exclude quasars with $J>20$ mag or with BH mass measurement uncertainties $>0.5$ dex. These quasars have low spectral S\slash N, peculiar continuum shapes that are likely caused by intrinsic reddening, or significantly affected by strong telluric line residuals. Since all spectra were calibrated using available $J$-band magnitudes \citep{Shen2019GNIRS}, we also exclude one object, J1545+6028, due to its absence of $J$-band data. The remaining 34 quasars provide high completeness in BH mass estimates for the SDSS luminosity samples. The BH mass completeness is 20\slash 24, 9\slash 10 and 5\slash 13 in the SDSS main, overlap, and Stripe 82 samples (refereed to as the SDSS\_M, SDSS\_O, and SDSS\_S82 samples hereafter). These quasars are summarized in \autoref{tab:bhm-sample}. The BH mass completeness is $>80\%$ in the combined SDSS\_M + SDSS\_O sample. In this work, we use this combined sample (referred to as the SDSS\_MO sample hereafter) to constrain the active BHMF and ERDF for $z\sim 6$ quasars.

   \begin{figure*}
   \plottwo{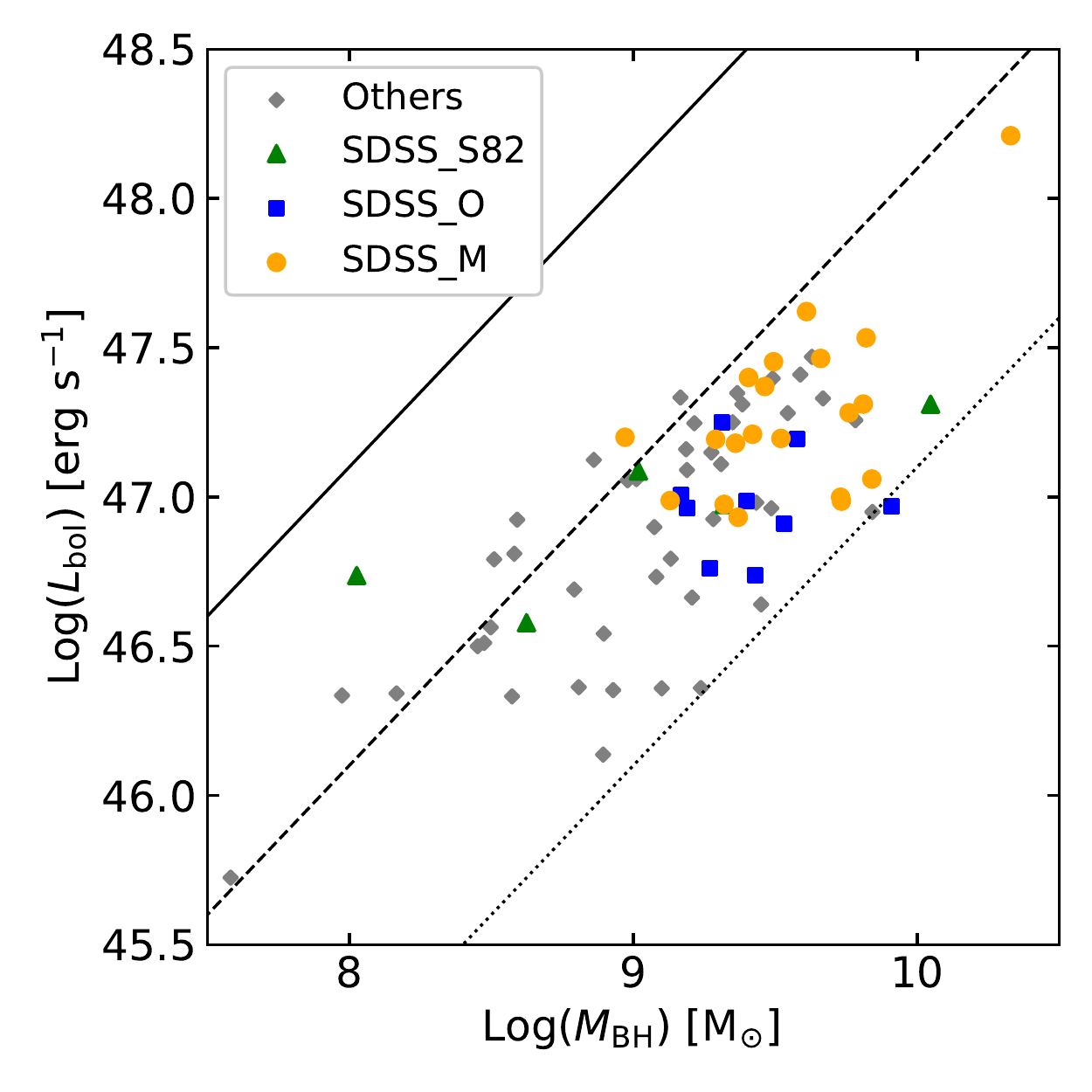}{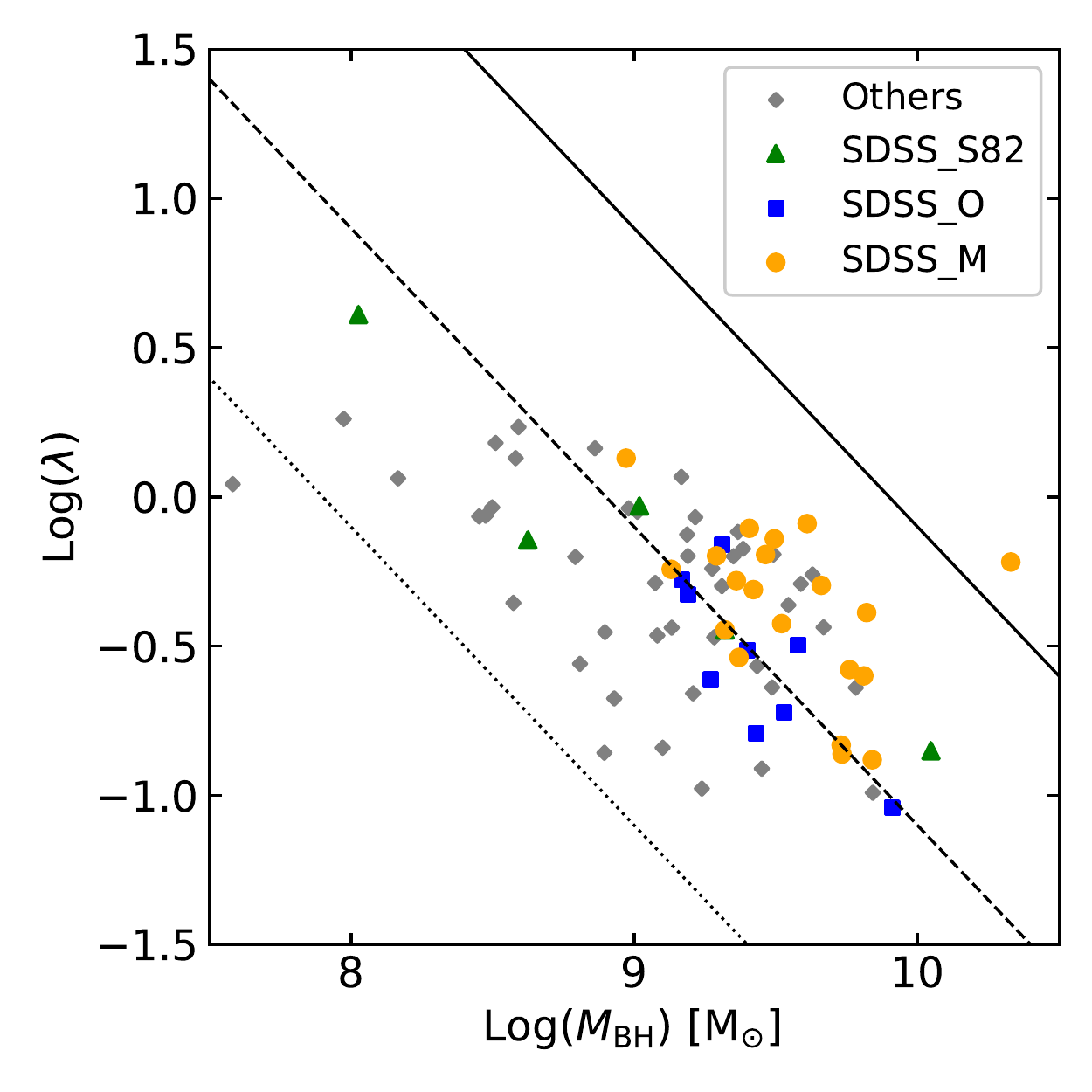}
   \caption{Observed distributions for the known quasars at $5.7 \leqslant z \leqslant 6.5$ in the BH mass-luminosity plane (left panel) and BH mass-Eddington ratio plane (right panel). The SDSS\_M, SDSS\_O, and SDSS\_S82 samples are shown in orange circle, blue square, and green triangle. The gray 	
diamonds are quasars from other surveys. The black solid, dashed, and dotted lines on the left panel indicate Eddington ratios of 10, 1, and 0.1, respectively. The black dotted, dashed and solid lines on the right panel indicate the bolometric luminosity of $10^{46}$, $10^{47}$, and $10^{48}$ $\mathrm{erg~s}^{-1}$, respectively.
Quasars in surveys other than SDSS are compiled from the literature \citep{Willott2010BHMF, Onoue2019, Eilers2020, Schindler2020}. Their virial masses are recalculated using the same mass recipes of this paper.
   }
   \label{fig:sample-2d}
   \end{figure*}

\subsection{Distributions of $M_\bullet$, $L_{\mathrm{bol}}$, and $\lambda_{Edd}$}
\label{distributionsofm_bulletl_mathrmbolandlambda_edd}

In \autoref{fig:sample-2d}, we show the bivariate distribution of the currently known quasars at $5.7 \leqslant z \leqslant 6.5$ in the BH mass-luminosity plane and mass-Eddington ratio plane. The BH masses of most quasars span a range from $\sim 10^8$ to $\sim 10^{10} M_\odot$ and the bolometric luminosities are in the range of $10^{46.1}$ to $10^{47.6} ~\mathrm{erg~s}^{-1}$.
There is one ultra-luminous quasar J0100+2802 with $M_{\mathrm{BH}}>10^{10} M_\odot$ and $L_{\mathrm{bol}} > 10^{48} ~\mathrm{erg~s}^{-1}$ reported by \citet{Wu15QSO}. There is another quasar J0859+0022 with a very low bolometric luminosity and a small BH mass from the SHELLQs survey \citep{Onoue2019}.
The SDSS samples cover about one third of the full BH mass sample for quasars at $5.7 \leqslant z \leqslant 6.5$. Compared with the full sample, the SDSS\_MO quasars on average have higher luminosities and higher BH masses. The SDSS\_S82 sample spans a larger BH mass range than SDSS\_MO, but it is too small and highly incomplete in terms of BH mass measurements.
 
For the SDSS quasars, there is a lack of objects in the low luminosity, low BH mass, and low Eddington ratio regions (lower-left corner in the $M_{\mathrm{BH}}-\lambda$ figure). These distributions are not caused by underlying distribution functions. Instead, they are affected by the specific survey selection criteria, i.e., we are only able to observe targets in certain mass and luminosity regions in different surveys. In contrast, the object density decline at the highest BH masses and Eddington ratios (upper-right corner in the $M_{\mathrm{BH}}-\lambda$ figure) is physical, caused by the decrease of the abundance in both BHMF and ERDF in this parameter regime. We will further illustrate these selection effects in our forward modeling results.
 
\begin{figure*}
   \begin{center}
    \includegraphics[keepaspectratio,width=0.95\textwidth, height=0.75\textheight]{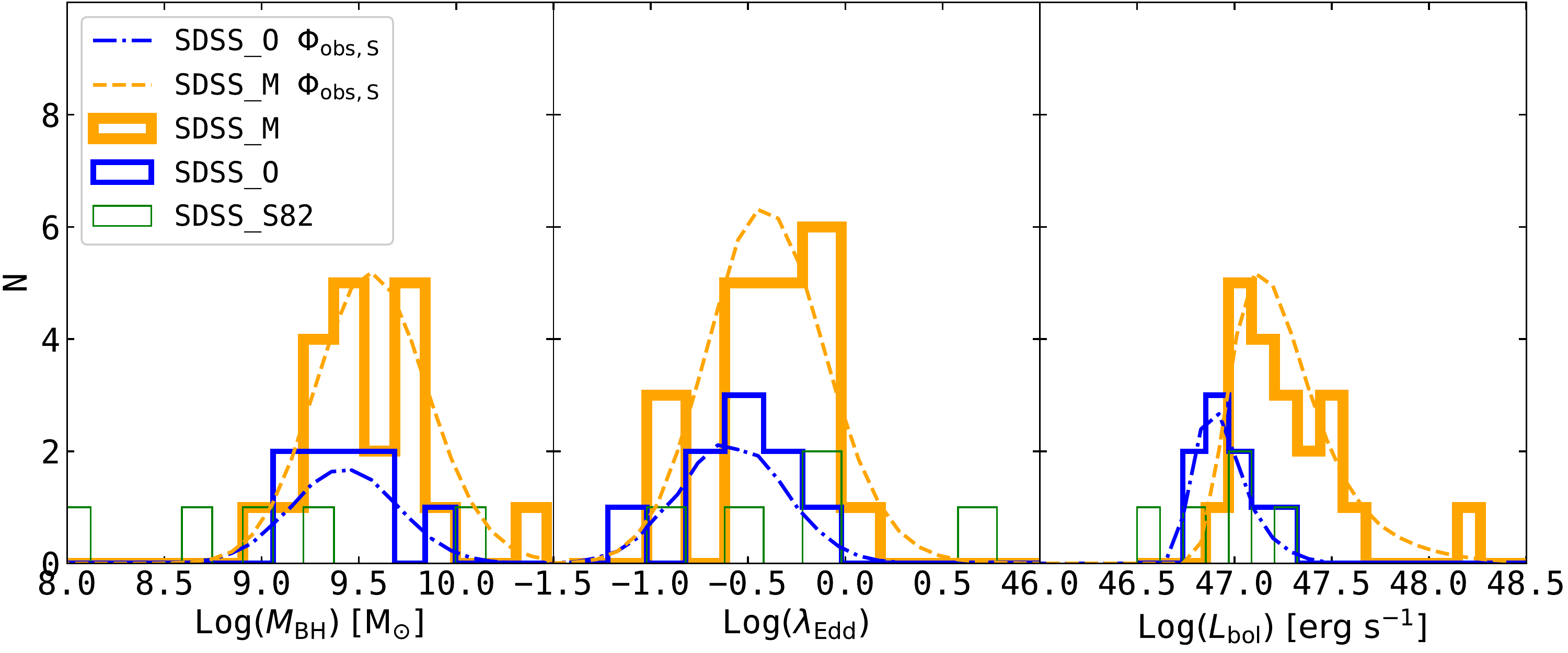}
   \caption{
   Histograms of BH masses (left panel), Eddington ratios (middle panel), and bolometric luminosities (right panel) for the SDSS samples. The orange, blue, and green bars represent the histogram data from SDSS\_M, SDSS\_O, and SDSS\_S82, respectively. The lines with the same colors are the predicted sample distributions from the fiducial models (see the definitions of \emph{survey observed} distributions in \autoref{sec:likelihood-method}).
   }
   \label{fig:bhmf-hist}
   \end{center}
   \end{figure*}

In \autoref{fig:bhmf-hist}, we show the 1D histograms of BH masses, Eddington ratios, and bolometric luminosities for the SDSS quasars. They are the 1D projections of \autoref{fig:sample-2d} and are also affected by the selection functions of different samples. The SDSS\_MO sample spans a mass range from $\sim 10^9$ to $\sim 10^{10.5} M_\odot$ and an Eddington ratio range from $\sim -1$ to $\sim 0$ in log$_{10}$ space. All quasars in our samples have Eddington ratios $\gtrsim 0.1$ and they are selected as point sources in each survey. The virial mass estimates are only available for bright, broad-line quasars, and thus our sample does not contain obscured (type 2) quasars. We will use this SDSS\_MO BH mass sample (29 objects) to jointly constrain the 2D demographics in the BH mass-luminosity plane (\S\ref{sec:likelihood-method} and \S\ref{sec:sdss-mo-results}), and then use the full luminosity sample (110 objects) to fit the BHMF only (\S\ref{sec:qlf-fitting-method} and \S\ref{sec:lf-results}).

\section{The Active BHMF and ERDF}
\label{sec:method}

There are several different methods to estimate the active BHMF in the literature. The $1/V_{\rm max}$ method directly uses the volume weights to measure binned BHMFs, as how QLFs are computed ~\citep[e.g.,][]{GreeneHo2007BHMF, Vestergaard2008BHMF}. It is difficult for this approach to properly account for sample incompleteness of the BHMF (especially at the low-mass end), due to the fixed flux limit and the luminosity range (Eddington ratio dispersion) in each mass bin ~\citep{Kelly2009BHMF}. This method also does not consider the uncertainties in the BH mass estimates.
For completeness, we will present the BHMF results using the $1/V_{\rm max}$ method as well, but we caution on the limitations of these $1/V_{\rm max}$-based results.
 
 \citet{Kelly2009BHMF} developed a forward-modeling Bayesian framework to estimate quasar BHMFs in a more rigorous manner.
 They built error models for directly measured quantities ($L_\lambda$, $\mathbf{v}$, z), where $L_\lambda$ and $\mathbf{v}$ are the quasar continuum luminosity and width of the broad emission lines used to calculate the virial mass. They used a mixture of Gaussian functions to describe the intrinsic BH mass-Eddington ratio distribution functions and probability distributions of physical quantities, e.g., $\rho(L_\lambda|M_\mathrm{BH})$ and $\rho(\mathbf{v}|L_\lambda, M_\mathrm{BH})$. This sophisticated model was applied to the SDSS quasar sample at $0.4<z<5$ ~\citep{ShenKelly2012BHMF,KellyShen2013BHMF}. The model had more than 10 parameters and was appropriate for large quasar samples.

 \citet{Schulze2010BHMF} developed a maximum likelihood method (also a forward modeling approach) to estimate the intrinsic BHMF and ERDF simultaneously.
 They assumed a joint distribution function of BH mass and Eddington ratio, fit the model parameters by maximizing the likelihood function, and normalized the distribution function by matching the predicted sample to the observed sample.  \citet{Schulze2015BHMF} updated the model to include uncertainties in virial BH masses. This approach has been applied to quasar samples at $1\lesssim z \lesssim 2$ ~\citep[e.g.,][]{Nobuta2012BHMF}.

Finally, an alternative approach to estimate the BHMF is to use the QLF, assuming a fixed Eddington ratio distribution model. This approach does not require BH mass estimates of quasars. \citet{Willott2010BHMF} derived the first $z\sim 6$ quasar BHMF using this approach. They adopted an observed Eddington ratio distribution from 17 $z\sim 6$ quasars known at that time, and obtained the BHMF from the QLF. This approach utilizes the often much larger sample of quasars with luminosity (but no BH mass) measurements. In this method, a priori knowledge of the Eddington ratio distribution is required, rather than constrained from the same sample as in the forward modeling approach.

In this paper, we will use the $1/V_{\rm max}$ method, the maximum likelihood approach (with modifications) and the QLF fitting approach for our BH mass sample and luminosity samples. Since our samples are small, the more sophisticated models in \citet{ShenKelly2012BHMF} and \citet{KellyShen2013BHMF} are not suitable. 

\subsection{The $1/V_{\rm max}$ Method}
\label{the1v_maxmethod}

Following common practice, we use the $1/V_{\mathrm{max}}$ method ~\citep{Avni1980VMAX} to estimate the QLF and active BHMF. $1/V_{\rm max}$ is the maximum accessible volume in which an object with a given magnitude can be selected, given the flux limit of the survey and the redshift bin used,
\[ V_{\rm max} = \frac{A}{4\pi} \int^{z_{\mathrm{\rm max}}}_{z_{\mathrm{\rm min}}} \Omega(L,z) \frac{\mathrm{d}V}{\mathrm{d}z} \mathrm{d}z\ , \] 
where $A$ is the sky coverage of the sample, $\mathrm{d}V/\mathrm{d}z$ is the differential comoving volume, $z_{\rm min}$ and $z_{\rm max}$ are the minimum and maximum redshifts for a quasar with luminosity $L$ that can be selected in the sample, and $\Omega(L,z)$ is the luminosity selection function mapped onto a 2D grid of luminosity and redshift. 

The binned BHMF is then
\[ \Phi(M_\bullet)=
   \frac{1}{\Delta \log M_\bullet} \sum^{N}_{j=1}(\frac{1}{V_{\mathrm{\rm max},j}}), \] 
with a Poisson statistical uncertainty
\[ \sigma(\Phi(M_\bullet))=
   \frac{1}{\Delta \log M_\bullet} [\sum^{N}_{j=1}(\frac{1}{V_{\mathrm{\rm max},j}})^2]^{1/2}\ , \] 
where the summation is over all quasars within a redshift-luminosity bin. Not all quasars in the luminosity sample have BH mass estimates, and we simply correct for this incompleteness by multiplying the fraction of those with mass estimates in the sample. For example, for the SDSS\_M sample, we multiply the result by a factor of 24/20, where we have 20 BH mass estimates out of 24 quasars in this sample. This simple correction assumes that quasars without BH mass estimates have on average the same mass distribution as those with mass estimates. Because the fraction of quasars with BH mass estimates is high among our designated BH mass samples, details in this correrction do not impact our final results. These numbers of quasars are listed in \autoref{tab:all-samples} for the relevant samples.

Similarly, we calculate the binned EDRF with the $1/V_{\rm max}$ method,
\[ \Phi(\lambda)=
   \frac{1}{\Delta \log \lambda} \sum^{N}_{j=1}(\frac{1}{V_{\mathrm{\rm max},j}}), \] 
\[ \sigma(\Phi(\lambda))=
   \frac{1}{\Delta \log \lambda} [\sum^{N}_{j=1}(\frac{1}{V_{\mathrm{\rm max},j}})^2]^{1/2}. \] 

The $1/V_{\rm max}$ approach does not account for virial BH mass uncertainties and does not properly account for selection incompleteness in terms of BH mass (since the selection is based on flux, not BH mass). Nevertheless, it provides a non-parametric and model-independent estimate of the BHMF. In this work, we only use the $1/V_{\rm max}$ results as a reference to compare with the \emph{measured} distributions (see \S\ref{sec:likelihood-method}) from the more rigorous forward modeling.

\subsection{The Maximum Likelihood Method}
\label{sec:likelihood-method}

\subsubsection{Basic Formalism}

We adopt a similar maximum likelihood approach as in \citet{Schulze2015BHMF}.
 This approach starts from a joint distribution function of BH mass and Eddington ratio $\Psi(M_\bullet, \lambda, z)$. Here, $\Psi(M_\bullet, \lambda, z) ~ \mathrm{d\,log}M_\bullet ~\mathrm{d\,log} \lambda$ describes the space density of quasars with masses between $\mathrm{log}M_\bullet$ and $\log M_\bullet + \mathrm{d\,log}M_\bullet$, and Eddington ratios between $\mathrm{log}\lambda$ and $\log\lambda + \mathrm{d\,log}\lambda$ at redshift $z$.
 We refer to it as the \emph{intrinsic} distribution function hereafter. This joint distribution function is equivalent to the joint distribution function in the mass-luminosity plane, i.e., $\Psi(M_\bullet, L_{\mathrm{bol}}, z)$, after changing the argument $\lambda$ to $L_{\rm bol}$ accordingly.

Given this joint distribution function, we can derive the 1D BHMF, ERDF, and QLF by integrating over other variables, i.e.,
   \begin{eqnarray}
   \Phi(M_\bullet, z) &=& \int \Psi(M_\bullet, \lambda, z) \mathrm{d\,log}\lambda\ ,   \label{eq:bhmf:intrinsic} \\
   \Phi(\lambda, z)   &=& \int \Psi(M_\bullet, \lambda, z) \mathrm{d\,log}M_\bullet\ , \label{eq:erdf:intrinsic} \\
   \Phi(L_{\rm bol}, z)   &=& \int \Psi(\lambda, L_{\rm bol}, z) \mathrm{d\,log}\lambda\ . \label{eq:lf:intrinsic}
   \end{eqnarray}
   \autoref{eq:bhmf:intrinsic}, \autoref{eq:erdf:intrinsic}, and \autoref{eq:lf:intrinsic} refer to the \emph{intrinsic} BHMF, ERDF, and QLF, respectively. 

The {\em measured} distribution (that is, the distribution based on the estimated quantities, before being modified by the selection function) is the convolution of the intrinsic distribution with the error model $g(M_{\bullet,e}, \lambda_e|M_\bullet,\lambda)$,
   \begin{eqnarray}
   \Psi_{err}(M_{\bullet,e}, \lambda_e, z) &=& \iint g(M_{\bullet,e}, \lambda_e|M_\bullet,\lambda) \nonumber \\
                                       & & \times \Psi(M_\bullet, \lambda, z) \mathrm{d\,log}M_\bullet~\mathrm{d\,log}\lambda\ ,
                                                 \label{eq:bhmf:err}
   \end{eqnarray}
 where subscript ``e'' is used to denote the \emph{measured} quantities (with errors in BH mass and bolometric luminosity estimates). We refer to $\Psi_{err}(M_{\bullet,e}, \lambda_e, z)$ as the \emph{measured} joint distribution function. 

The \emph{measured} BHMF, ERDF, and QLF are then
   \begin{eqnarray}
   \Phi_{err}(M_{\bullet,e}, z) &=& \int \Psi_{err}(M_{\bullet,e}, \lambda_{e}, z) \mathrm{d\,log}\lambda\ , \label{eq:bhmf:measured} \\
   \Phi_{err}(\lambda_{e}, z)   &=& \int \Psi_{err}(M_{\bullet,e}, \lambda_{e}, z) \mathrm{d\,log}M_\bullet\ , \label{eq:lambda:measured} \\
   \Phi_{err}(L_{\mathrm{bol},e}, z)   &=& \int \Psi_{err}(\lambda, L_{\mathrm{bol},e}, z) \mathrm{d\,log}\lambda\ . \label{eq:lf:measured}
   \end{eqnarray}

   
Finally, the \emph{observed} distribution, $\Psi_{obs}(M_{\bullet,e}, \lambda_e, z)$, is the measured distribution modified by the sample selection functions, $\Omega(M_\bullet,  \lambda, z)$, or $\Omega(L_{\mathrm{bol}}, z)$, depending on the distribution in question. Hereafter we will omit the subscript ``e'' in the input arguments of $\Psi_{err}$ and $\Psi_{obs}$; they are still the distributions of the estimated quantities in \autoref{eq:bhmf:err}. Therefore we obtain
\[ \Psi_{obs}(M_\bullet, \lambda, z) = \Omega(M_\bullet,  \lambda, z) \Psi_{err}(M_\bullet, \lambda, z) \addtag \label{eq:psi:obs}\ , \]
which will be compared with the observed sample to constrain the model parameters. 

The observed BHMF, ERDF, and QLF are derived similarly as in \autoref{eq:bhmf:intrinsic}, \autoref{eq:erdf:intrinsic}, and \autoref{eq:lf:intrinsic}, by substituting $\Psi(M_{{\mathrm{BH}}}, \lambda, z)$ with $\Psi_{obs}(M_{{\mathrm{BH}},e}, \lambda_e, z)$.
If we further marginalize over the redshift range of the sample, we obtain \emph{observed} BHMF (\autoref{eq:bhmf:obs}), ERDF (\autoref{eq:erdf:obs}), and QLF (\autoref{eq:lf:obs}) of our $z\sim 6$ sample (space densities in units of $\mathrm{Mpc}^{-3} \mathrm{dex}^{-1}$). Integrating over the observed volume, we obtain the \emph{survey observed} distributions in units of counts (\autoref{eq:bhmf:obs_survey} $\sim$ \autoref{eq:lf:obs_survey}), which will need to match the distributions of data (as shown in \autoref{fig:bhmf-hist}):
   \begin{eqnarray}
   \Phi_{obs}(M_\bullet) &=& \iint \Psi(M_\bullet, \lambda, z) \mathrm{d\,log}\lambda   ~ \mathrm{d\,}z \label{eq:bhmf:obs} \\
   \Phi_{obs}(\lambda)   &=& \iint \Psi(M_\bullet, \lambda, z) \mathrm{d\,log}M_\bullet ~ \mathrm{d\,}z \label{eq:erdf:obs} \\
   \Phi_{obs}(L_{\mathrm{bol}})   &=& \iint \Psi(\lambda, L_{\mathrm{bol}}, z) \mathrm{d\,log}\lambda ~ \mathrm{d\,}z \label{eq:lf:obs} \\
    \Phi_{obs,S}(M_\bullet) &=& A \iint \Psi(M_\bullet, \lambda, z) \frac{\mathrm{d}\,V}{\mathrm{d}\,z} \mathrm{d\,log}\lambda   ~ \mathrm{d\,}z \label{eq:bhmf:obs_survey} \\
   \Phi_{obs,S}(\lambda)   &=& A \iint \Psi(M_\bullet, \lambda, z)  \frac{\mathrm{d}\,V}{\mathrm{d}\,z} \mathrm{d\,log}M_\bullet ~ \mathrm{d\,}z \label{eq:erdf:obs_survey} \\
   \Phi_{obs,S}(L_{\mathrm{bol}})   &=& A \iint \Psi(\lambda, L_{\mathrm{bol}}, z)  \frac{\mathrm{d}\,V}{\mathrm{d}\,z} \mathrm{d\,log}\lambda ~ \mathrm{d\,}z \label{eq:lf:obs_survey}
   \end{eqnarray}
where $A$ is the effective area of the survey and ${\mathrm{d}\,V}/{\mathrm{d}\,z}$ is the differential volume.
   
The maximum likelihood method aims at minimizing the likelihood function $S = -2 \mathrm{ln} \mathcal{L}$, where the total likelihood $\mathcal{L} = \prod^N_{i=1} p_i$ is the product of the individual likelihoods for the observed objects, and $p_i(M_{{\mathrm{\bullet}}}, \lambda, z)$ is given by the normalized \emph{observed} distribution
   \begin{eqnarray}
   p_i(M_\bullet, \lambda, z) =& \frac{1}{N_i} \Psi_{obs}(M_\bullet, \lambda, z)      \frac{\mathrm{d}\,V}{\mathrm{d}\,z} \nonumber \\
   = &\frac{1}{N_i} \Omega_i(M_\bullet,  \lambda, z) \Psi_{err}(M_\bullet, \lambda, z)\frac{\mathrm{d}\,V}{\mathrm{d}\,z} \nonumber \ ,
   \end{eqnarray}
where $\Omega_i(M_\bullet, \lambda, z)$ is the selection function for the \emph{i}th object and 
   \begin{eqnarray}
   N_i = A_i \iiint &\Omega_i(M_\bullet,  \lambda, z) \Psi_{err}(M_\bullet, \lambda, z)  \nonumber \\
               &\frac{\mathrm{d}\,V}{\mathrm{d}\,z}~\mathrm{d\,log}M_\bullet~\mathrm{d\,log}\lambda~\mathrm{d\,}z \label{eq:bhmf:N}
   \end{eqnarray}
is the normalization for the \emph{i}th object. If all objects are in the same survey, the selection function will be the same, and $N_i$ will be the \emph{observed} number of objects for that specific survey. Otherwise, individual likelihoods are calculated separately for each survey in the sample.

We then minimize the likelihood function
\[ S = -2 \sum^N_{i=1}\left [ \ln \Omega_i(M_\bullet,  \lambda, z) \Psi_{err}(M_\bullet, \lambda, z) - \ln N_i \right ]. \] 

Following \citet{Schulze2015BHMF}, this maximum likelihood method adopts a forward modeling approach. It generates the \emph{observed} distribution function and finds the best parametric \emph{intrinsic} distribution function by matching data in the $M_\bullet-\lambda$ plane. This approach only uses the shape of the distribution to constrain the underlying \emph{intrinsic} distribution function. The normalization of $\Psi(M_\bullet, \lambda, z)$ is determined by integrating over the model and scaling the predicted number of objects to the observed number in our sample via
\[ \Psi_* = \frac{\sum_j N_{data,j}}{\sum_j N_{model,j}}\ , \addtag \label{eq:bhmf:normal} \]
where $N_{data,j}$ is the number of observed quasars and $N_{model,j}$ is the predicted number from \autoref{eq:bhmf:N}. The summation is over all surveys denoted by index \emph{j}.
There are two caveats when using \autoref{eq:bhmf:normal} directly for our fiducial BH mass sample (i.e., the SDSS\_MO=SDSS\_M+SDSS\_O sample).
First, all surveys should be statistically independent and have no overlap in the luminosity-sky coverage space. While the SDSS\_O and SDSS\_M samples do have some overlap in sky coverage, we have separated both samples in luminosity in constructing the selection functions \citep{Jiang2016SDSSQ}, thus they satisfy the above requirement.
Second, the normalization factor calculated by \autoref{eq:bhmf:normal} is averaged over all surveys. Since our SDSS\_MO sample is dominated by the SDSS\_M sample, we choose to normalize the distribution function using the SDSS\_M sample size only, to avoid statistical fluctuations from the smaller SDSS\_O sample. 

 
Compared with earlier BHMF studies at lower redshift ~\citep[e.g.][]{ShenKelly2012BHMF, Nobuta2012BHMF, Schulze2015BHMF}, our sample has one additional complication in luminosity conversion. In the work by \citet{ShenKelly2012BHMF}, the bolometric luminosity is computed from the $i$-band magnitude and the survey selection function is in the form of $\Omega(m_i, z)$.  \citet{Nobuta2012BHMF} calculated the selection function from their spectroscopic data and obtained $\Omega(M_\bullet, \lambda, z)$ directly. These samples have a direct mapping of the selection function to the luminosity-redshift space, which largely simplifies the modeling of the BHMF and ERDF.
 For our $z\sim 6$ quasars, the $M_{1450}$ magnitude were measured from their discovery optical spectra and $z$-band magnitudes, and the selection function is defined in terms of $\Omega(M_{1450}, z)$. To perform our forward modeling, we require the selection function defined in the $L_{\rm bol}$-redshift space. 
  
   \begin{figure}
   \plotone{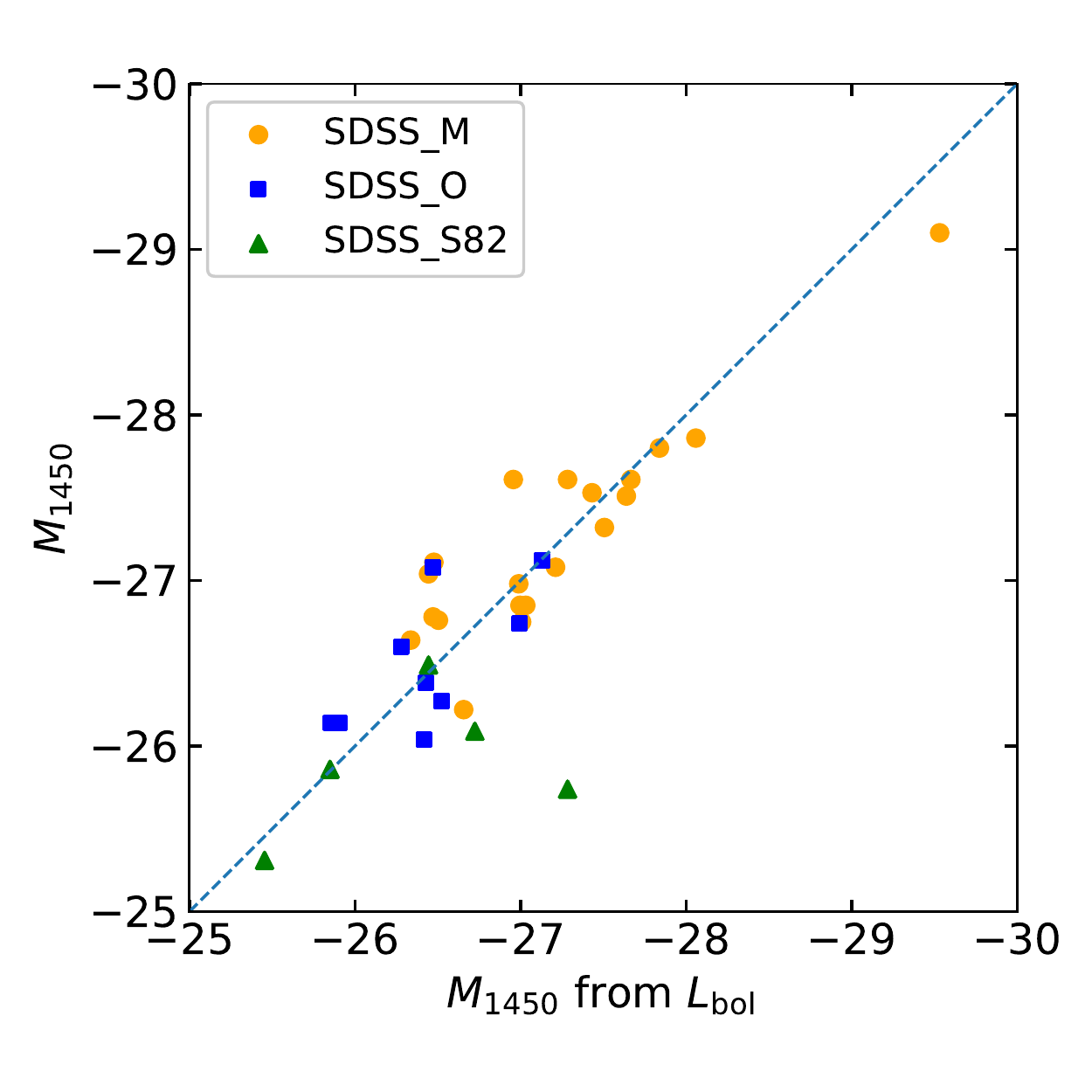}
   \caption{Scatter in $M_{1450}$ measured from different methods. The X-axis represents $M_{1450}$ converted from $L_{\mathrm{bol}}$ after the offset correction. The Y-axis corresponds to the $M_{1450}$ measured from the optical spectra and the $z$-band magnitudes \citep{Jiang2016SDSSQ}. The scatter is about 0.35 mag. See \autoref{sec:likelihood-method} for details. \label{fig:c-correct}}
   \end{figure}

We convert $L_{\mathrm{bol}}$ to $L_{3000}$ using a bolometric correction factor of 5.15 ~\citep{Richards2006SED}, and then convert $L_{3000}$ to $M_{1450}$ assuming a power-law continuum spectrum $f_\lambda \propto \lambda^{-1.5}$. Compared with the $M_{1450}$ measured from the optical spectra, there are both scatter and offset in our re-derived $M_{1450}$.
A similar scatter is also observed in the CFHQS sample by \citet{Matsuoka2018QLF}. The $M_{1450}$ magnitudes of the CFHQS quasars were originally estimated from the observed $J$-band fluxes with a template quasar spectrum. For consistency with the measurements in SDSS and SHELLQs, \citet{Matsuoka2018QLF} re-measured $M_{1450}$ by extrapolating the continuum spectrum redward of \lya, assuming a power-law form $f_\lambda \propto \lambda^{-1.5}$. The calculated $M_{1450}$ values differ from the original (CFHQS) values by $-0.4$ to 0.2 mag, including one faint quasar with an offset of $-0.7$ mag. These differences could be from the scatter in the continuum slope of different quasars. The measurement uncertainties of $M_{1450}$ and $L_{\mathrm{bol}}$ may also contribute to the scatter. In this work, we assume a Gaussian distribution for $M_{1450}$ at fixed $L_{\mathrm{bol}}$. The scatter is 0.35 mag, or 0.14 dex in luminosity. This extra scatter in $L_{\rm bol}-M_{1450}$ conversion is incorporated in our error model.

In addition to the scatter between $M_{1450}$ and $L_{\rm bol}$, we also find an offset in the conversion. Assuming a power-law spectrum $f_\lambda \propto \lambda^{-1.5}$ and a bolometric correction of 5.15 for $L_{3000}$, the bolometric correction from $L_{1450}$ to $L_{\mathrm{bol}}$ should be 3.6. We find that the converted $M_{1450}$ is on average fainter than the measured $M_{1450}$ by 0.27 mag. We correct this offset in the conversion process to match the QLF of our BH mass sample with the QLF in \citet{Matsuoka2018QLF}, which is shown in the right panel of \autoref{fig:bhmf-1d} (see further discussion in \autoref{sec:sdss-mo-results}).
After correcting for the average offset, the comparison between the $M_{1450}$ measured from the optical spectra and the $M_{1450}$ converted from $L_{\mathrm{bol}}$ is shown in \autoref{fig:c-correct}.

Now we can define our final error model $g(M_{\bullet,e}, \lambda_e|M_{\bullet},\lambda)$ in \autoref{eq:bhmf:err}:
    \begin{eqnarray}
    &&g(M_{\bullet,e}, \lambda_e|M_{\bullet},\lambda) = g(m_e, l_e|m,l) = \frac{1}{2\pi\sigma_{\mathrm{VM}} \sigma_{\mathrm{bol}}} \times  \nonumber \\
    &&\exp \lbrace -\frac{(m_e-(m + \beta_e(l - \hat{l}(m)) ))^2}{2\sigma_{\mathrm{VM}}^2} - \frac{(l_e-l)^2}{2\sigma_{\mathrm{bol}}^2} \rbrace \label{eq:error-model} \\
    &&\hat{l} (m)= \log_{10}(1.26 \times 10^{38}) + m + \int \log \lambda ~ \rho_\lambda(\lambda, m)~\mathrm{d\,log}\lambda\ ,  \nonumber
    \end{eqnarray}
    where $m = \log M_\bullet$, $l = \log L_{\mathrm{bol}}$, $\sigma_{\mathrm{VM}}$ denotes the scatter of virial mass estimates at a fixed true mass and a fixed luminosity, the error slope $\beta_e$ describes the level of mass bias in the measured virial mass at a fixed true mass and luminosity, and $\hat{l} (m)$ is the expectation value of luminosity at a fixed true mass, determined by the Eddington distribution. Since the luminosity uncertainties are random scatter around the $R-L$ relation, we use $\beta_e=0.2$
    (the best-fitting value when we set $\beta_e$ as a free parameter; see more discussion in \autoref{sec:errors}).

Most studies have neglected a potential luminosity-dependent bias in single-epoch virial masses by assuming $\beta_{e}=0$. The empirical virial mass recipes are calibrated for average luminosities of quasars at each fixed mass. At a fixed true mass, the instantaneous luminosity follows the Eddington ratio distribution. If the broad line width perfectly responds to the luminosity difference at the fixed true mass \citep[a behavior known as the broad-line region ``breathing'', e.g.,][]{Wang_breathing_2020}, then there is no luminosity-dependent bias in virial masses. However, if the broad line used for virial masses deviates from normal breathing \citep{Yang2020MgIIVariability,Wang_breathing_2020}, there will be a luminosity-dependent bias in virial masses, as emphasized in e.g., \citet{Shen2008BHBias, ShenKelly2012BHMF,Shen2013BHM}. Indeed, \citet{Guo_etal_2020b} compiled a large sample of quasars with multi-epoch spectra from the SDSS, and found that the virial masses (based on \CIV) are systematically higher in the bright state than in the faint state,  demonstrating the existence of this luminosity-dependent bias in virial masses. Following earlier work \citep[e.g.,][]{ShenKelly2012BHMF}, we incorporate a non-zero $\beta_{e}$ in our error model of \autoref{eq:error-model}.  

For the uncertainty in the bolometric luminosity $\sigma_{\mathrm{bol}}$, it mainly originates from the uncertainties in the bolometric correction and from the scatter in luminosity conversion as discussed earlier. The typical measurement uncertainty of $L_{\mathrm{bol}}$ is $\lesssim 0.05$ dex in our sample, hence is negligible. In this work we adopt $\sigma_{\mathrm{bol}}=0.14$ dex, which is the observed scatter from luminosity conversions (the uncertainty associated with the bolometric correction is ignored). This uncertainty in the bolomeric luminosity does not contribute to the uncertainty in the viral BH mass, because the latter is calculated directly from the monochromatic continuum luminosity. 

The systematic uncertainty associated with single-epoch viral masses is typically $\sim 0.4$ dex ~\citep[e.g.][]{Vestergaard_Peterson_2006, Shen2013BHM}, which is the dispersion in virial masses at a fixed true mass (averaged over instantaneous luminosities). \citet{KellyShen2013BHMF} adopted $\sigma_{\rm VM} \sim 0.4$ dex for their $\beta_e=0$ model. In our fiducial model, we adopt $\sigma_{\mathrm{VM}} = 0.25$ and $\beta_e=0.2$, which are close to the best fitting parameters when we set them as free parameters in the fitting process. More discussions about the choices of $\sigma_{\mathrm{VM}}$ and $\beta_e$ are presented in \autoref{sec:errors}. 

\subsubsection{Detailed Parameterization of Distribution Functions}
\label{parametricmodel}

   \begin{figure*}
   \plottwo{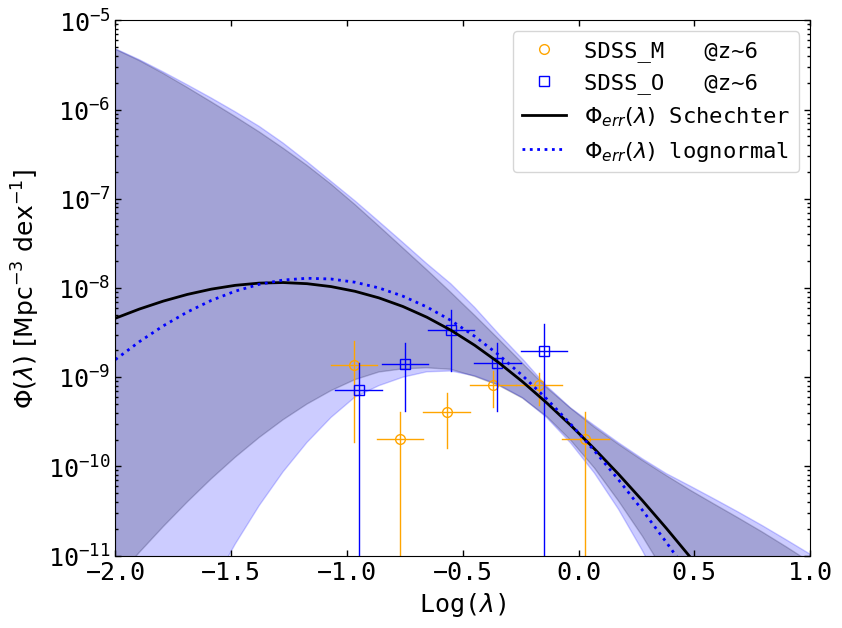}{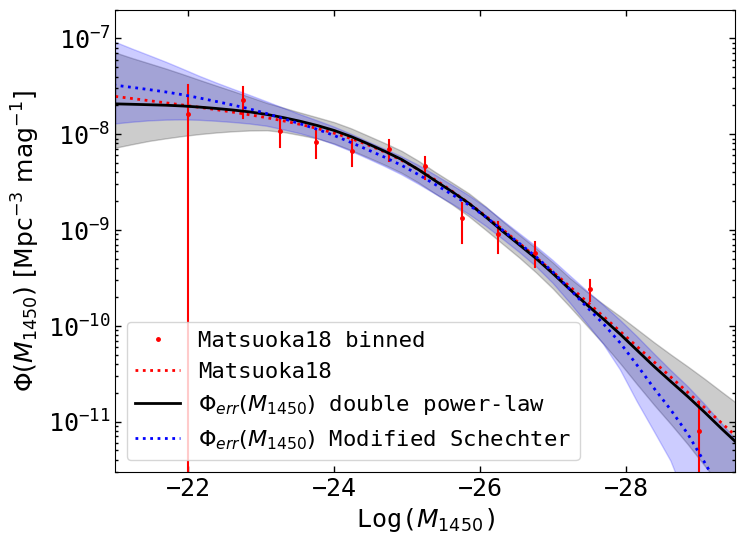}
   \caption{Fitting results of different model functions. The left panel shows the ERDF results of the maximum likelihood method, with a double power-law function as the mass term. The black solid and blue dotted lines represent the best fitting models of the measured ERDF, using the Schechter function and a lognormal distribution function as the Eddington ratio term. The gray and light blue shadow regions represent the $1\sigma$ regions of the two models. The orange circle and blue squares show the ERDF calculated by the $1/V_{\rm max}$ method for the SDSS\_M and SDSS\_O samples.
   The right panel shows the UV QLF of the QLF fitting method, with Schechter function as the Eddington ratio term. The black solid and blue dotted lines represent the best fitting model of the measured QLF, using a double power-law function and modified Schechter function as the mass term. The red points and dotted line represent the QLF result from \citet{Matsuoka2018QLF}. See \autoref{sec:model-compare} for details.
   }
   \label{fig:model-compare}
   \end{figure*}

For the maximum likelihood approach, we assume a parametric model for the joint distribution function $\Psi(M_\bullet, \lambda, z)$. Following  \citet{KellyShen2013BHMF} and \citet{Schulze2015BHMF}, we start from an intrinsic (active) BHMF and an ERDF that has a mass dependence. The intrinsic distribution function can be written as
\[ \Psi(M_\bullet, \lambda, z) = \rho_\bullet(M_\bullet) \rho_\lambda(\lambda, M_\bullet)\rho_z(z)\ , \addtag \label{eq:rho:full} \]
where $\rho_\bullet(M_\bullet)$ is the BH mass term, $\rho_\lambda(\lambda, M_\bullet)$ is the Eddington ratio term, and $\rho_z(z)$ is the redshift dependence term. The BHMF, ERDF, and QLF can then be derived from \autoref{eq:bhmf:intrinsic}, \autoref{eq:erdf:intrinsic}, and \autoref{eq:lf:intrinsic}, respectively.

For the mass term, we adopt the double power-law model, which is a simplification from the Gaussian mixture function adopted in \citet{kelly2012BHMF}, given our small sample size:
\[ \rho_\bullet(M_\bullet) = \frac{\Psi^*}{(M_\bullet/M^*_\bullet)^{-(\alpha+1)} + (M_\bullet/M^*_\bullet)^{-(\beta+1)}}\ , \] 
where $\Psi^*$ is the normalization factor, $\alpha$ and $\beta$ are the slopes at the low-mass and high-mass ends of the distribution, and $M^*_\bullet$ is the break mass scale.
We also test a modified Schechter function for the mass term ~\citep{Schulze2010BHMF},
\[ \rho_\bullet(M_\bullet)=\Psi^*(\frac{M_\bullet}{M_\bullet^*})^{\alpha+1}\exp{\left ( -\left [\frac{M_\bullet}{M_\bullet^*}\right ]^\beta \right )}. \] 

For the Eddington ratio term, we use the Schechter function ~\citep{Schechter1976} with a mass dependence,
\[ \rho_\lambda(\lambda, M_\bullet)=\left (\frac{\lambda}{\lambda_*(M_\bullet)}\right )^{\alpha_{\lambda}+1}\exp{\left ( -\frac{\lambda}{\lambda_*(M_\bullet)} \right )}  \addtag \label{eq:rho:lambda}\ , \]
where the mass dependence in $\lambda_*$ is
\[ \log \lambda_*(M_\bullet) = \log \lambda_0 + k_\lambda(\log M_\bullet - \log M_{\bullet, 0})\ . \addtag \label{eq:lambda-star} \]
We set the constant $\log M_{\bullet, 0}=9.5$ for our sample with a typical mass range of $\log M_{\bullet}=9\sim10$. We also test a lognormal model for the Eddington ratio term,
\[ \rho_\bullet(\lambda, M_\bullet) = \frac{1}{\sqrt{2\pi}\sigma_\lambda} \exp \left \{ -\frac{(\log \lambda - \log \lambda_*(M_\bullet))}{2\sigma_\lambda^2} \right \} \ , \] 
where $\lambda_*$ has the same parameterization as in \autoref{eq:lambda-star}, and $\sigma_{\lambda}$ is an extra parameter for the dispersion of Eddington ratios at fixed BH mass $M_{\bullet}$.

In \autoref{eq:rho:lambda}, $\lambda_*$ is the characteristic value of the Schechter function. When $\lambda$ is larger than $\lambda_*$, the exponential part in the equation becomes dominant and the density drops rapidly with $\lambda$. For the lognormal model, $\lambda_*$ has similar effects. We will use the term \emph{break point} to represent the $\log \lambda$ value when the density starts to drop rapidly for both the Schechter and the lognormal models.
The comparisons between different BHMF and ERDF models are presented in \autoref{sec:model-compare} and \autoref{fig:model-compare}.

We clarify that the mass term is not the BHMF, which is calculated through the integration in \autoref{eq:bhmf:intrinsic}.
When the Eddington ratio term is lognormal, the BHMF is proportional to the mass term. But with the general form of the Eddington ratio term, the shape of the BHMF is related to the Eddington ratio term as well as the integration range. This is also the case for the Eddington ratio term ($\rho_\lambda(\lambda, M_\bullet)$) and the actual ERDF.

The redshift evolution term is parameterized as follows,
\[ \rho_z(z) = 10^{k(z-z_0)}\ , \]
where $k$ is the evolution scale factor and has been well measured in high-redshift quasar QLF studies ~\citep[e.g.][]{Jiang2016SDSSQ, WangFeige2019b}. Here we adopt the value of $k=-0.7$ from  \citet{Jiang2016SDSSQ}. We also fixed $z_0=6$, i.e., the central point of our redshift range.

The absolute normalization of the intrinsic distribution function is obtained by scaling \autoref{eq:bhmf:N} to match the observed sample size of SDSS\_M.  In the fitting process, we use integration intervals of $-2.5<\log \lambda <0.5$, $6.0<\log M_\bullet<10.5$, and $5.7<z<6.5$. These integration ranges are sufficiently large for convergence and to cover plausible parameter ranges of $z\sim 6$ quasars. Our model has six free parameters $(M_*, \alpha, \beta, \lambda_x, \lambda_0, k_\lambda)$ to be determined in the fitting procedure, and $\Psi^*$ is determined in the normalization step. Here $\lambda_x$ is $\alpha_\lambda$ in the Schechter Eddington ratio term and $\sigma_{\lambda}$ in the lognormal Eddington ratio term.

\subsection{The QLF Fitting Method}
\label{sec:qlf-fitting-method}

In the maximum likelihood method, we fit the intrinsic distribution function using the BH mass sample in the BH mass-Eddington ratio plane.
 Among the nearly 300 quasars known at $5.7 \leqslant z \leqslant 6.5$, only $\sim$90 of them have BH mass measurements ~\citep[e.g.,][]{Willott2010BHMF, Jiang07BHM, DeRosa2011BHM, Wu15QSO, Mazzucchelli2017, Shen2019GNIRS, Onoue2019, Eilers2020, Schindler2020}. The SDSS\_M and SDSS\_O samples are the only flux-limited samples with mass completeness higher than 80\%. The total sample size of SDSS\_M+SDSS\_O is 29. As show in \autoref{fig:sample-2d} and \autoref{fig:bhmf-hist}, quasars in SDSS\_M and SDSS\_O have bolometric luminosities $\gtrsim 10^{46.5} \  \mathrm{erg~s}^{-1}$ and masses $\gtrsim 10^9 M_\odot$. As a result, the constraints on the low-mass end ($M_\bullet<10^9 M_\odot$) of the BHMF are poor.

Compared with the BH mass sample, the luminosity sample is much larger and extends to lower luminosities and lower masses.  \citet{Matsuoka2018QLF} combined $z\sim 6$ quasars in the SDSS, CFHQS, and SHELLQs surveys to derive reliable constraints on the QLF in the luminosity range of $-23<M_{1450}<-30$. Given a fixed $\rho_\lambda(\lambda, M_\bullet)$ model, we can calculate QLF from \autoref{eq:lf:measured}, and thus constrain the BHMF by matching the luminosity data in the observed luminosity sample.

To do so, we use the combined SDSS, CFHQS, and SHELLQs luminosity sample, and use the same $\rho_\lambda(\lambda, M_\bullet)$ model constrained from our maximum likelihood approach and the BH mass sample (\S\ref{sec:likelihood-method}). With this QLF-fitting method, we can achieve better constraints on the BHMF in the low-mass regime. 

For this purpose, we use the QLF likelihood function from  \citet{Marshall1983Likelihood} and minimize
   \begin{eqnarray}
   S = &-2 \sum^N_{i=1} [ \ln \Omega_i(M_i, z) \Phi_{err}(M_i, z)] \nonumber \\
       &+2 \iint \Omega_i(M_i, z) \Phi_{err}(M_i, z) \frac{\mathrm{d}\,V}{\mathrm{d}\,z} \mathrm{d\,}M~\mathrm{d\,}z\ ,
   \end{eqnarray}
where $M$ is the $M_{1450}$ magnitude, $\Omega_i(M_i, z)$ is the selection function, and $\Phi_{err}(M, z)$ is the QLF function converted from \autoref{eq:lf:measured}. This model has four free parameters $(\Psi^*, M_*, \alpha, \beta)$. The normalization factor $\Psi^*$ is directly constrained by this likelihood function and there is no additional normalization step.

\section{Results}
\label{sec:result}

There are two steps in our fitting procedure. First, we apply the maximum likelihood method to the SDSS\_MO sample, jointly constraining the distribution in the mass-luminosity plane (or equivalently, the mass-Eddington ratio plane). Then, we apply the QLF fitting method to the luminosity sample and obtain a better constrained BHMF in the low mass regime, during which the Eddington ratio distribution is fixed to the best-fitting model in the prior maximum likelihood approach. Our final results include the 2D distribution function in the mass-luminosity plane from the maximum likelihood step, as well as the improved BHMF results from the QLF fitting step.
 
\begin{figure*}
   \subfigure[BHMF]{\includegraphics[keepaspectratio,width=0.32\textwidth]{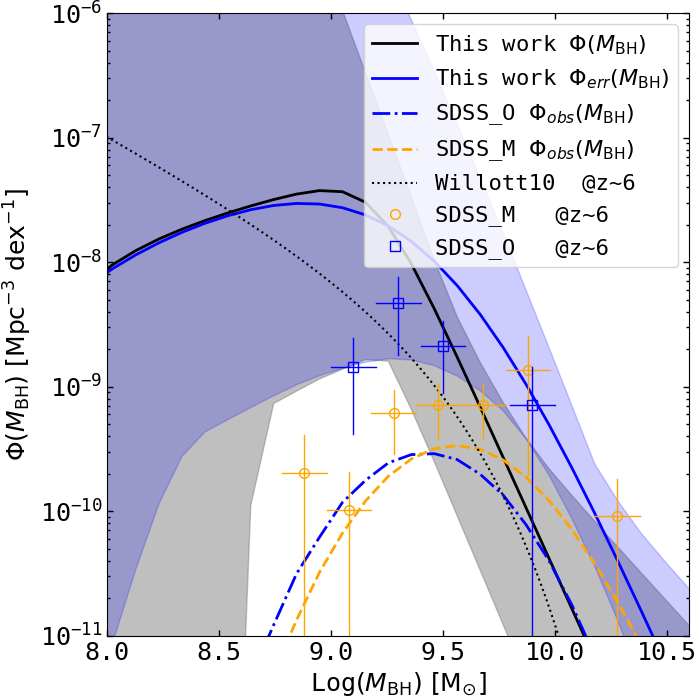}}
   \subfigure[ERDF]{\includegraphics[keepaspectratio,width=0.32\textwidth]{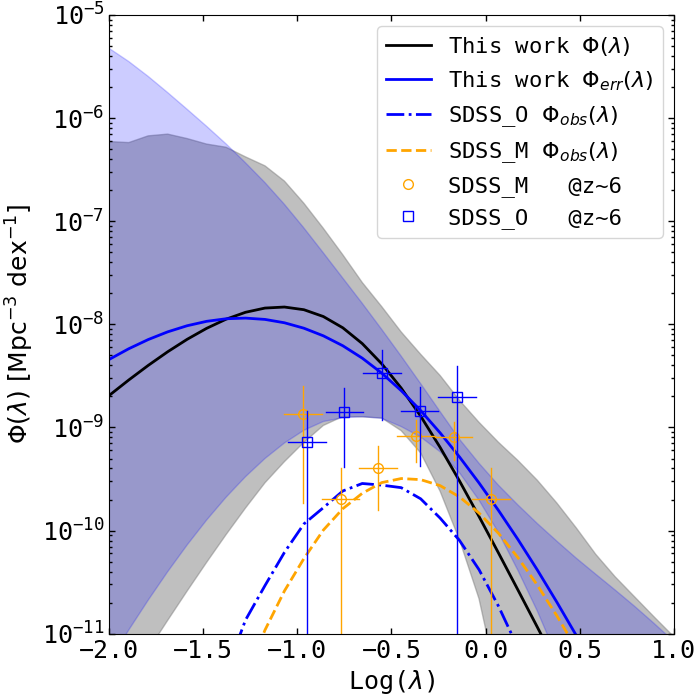}}
   \subfigure[QLF]{\includegraphics[keepaspectratio,width=0.32\textwidth]{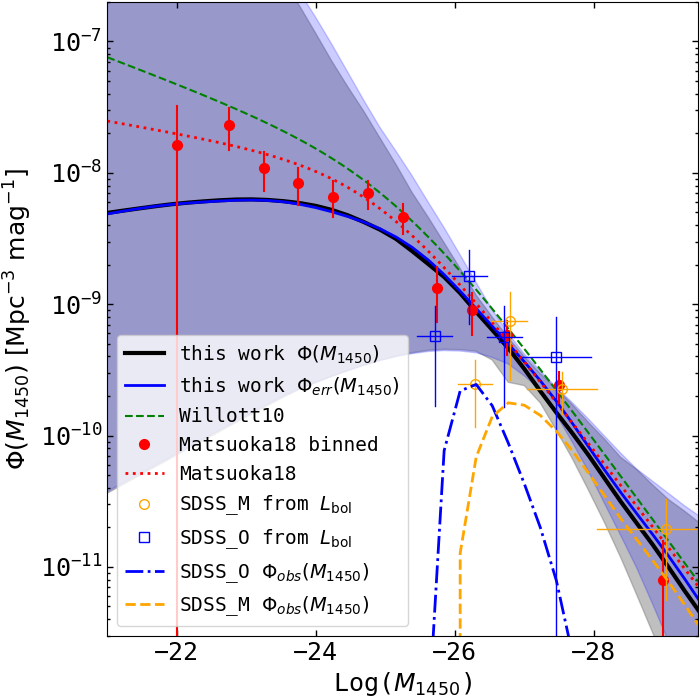}}
   \caption{Fitting results of the maximum likelihood method for the SDSS\_MO sample. The BHMF, ERDF, and QLF at $z=6$ are shown in the left, middle, and right panel, respectively. In the left panel, the black and blue solid lines are the intrinsic BHMF and measured BHMF of the best fitting model. The gray and light blue shadow regions represent their $1\sigma$ regions. The orange dashed and blue dash-dot lines show the observed BHMF in the SDSS\_M and SDSS\_O fields. The orange circle and blue squares with error bars are the BHMF calculated by the $1/V_{\rm max}$ method. These lines and symbols above have the same meanings in the ERDF and QLF panels as well. In the left panel, the black dotted line is the active BHMF from \citet{Willott2010BHMF}. In the right panel, the red dotted line and points show the $z=6$ QLF from \citet{Matsuoka2018QLF}. The green line represents the $z=6$ QLF from \citet{Willott2010QLF}. The orange circle and blue squares are the binned QLF of our sample (the $M_{1450}$ luminosity is calculated from bolometric luminosities).
   }
   
   \label{fig:bhmf-1d}
   \end{figure*}

From the maximum likelihood method, the 2D distribution in the BH mass-luminosity plane or the mass-Eddington ratio plane provides us the complete demographic information of $z\sim 6$ quasars, while the 1D BHMF, ERDF, and QLF are simply marginalized distributions of the 2D distribution. In some cases, the 1D integrated functions depend on the integration range. In short, the 1D distributions will have a much higher density for BHMF and QLF in the low mass (low luminosity) region if the integration is extended to very low mass or luminosity ranges that have not been explored observationally. In the main text, we use default integration ranges of $-2.5<\log \lambda <0.5$, $8.5<\log M_\bullet<10.5$, and $5.7<z<6.5$ to calculate these 1D functions. The effects of integration ranges are further discussed in \autoref{app:a}. For all figures with 1D distributions (e.g., \autoref{fig:bhmf-1d}), the $1\sigma$ uncertainty ranges (shaded area in these figures)
represent the range that corresponds to the top 68\% posterior probabilities of parameters in the Monte Carlo Markov chain.

\subsection{Fiducial Model Distributions}
\label{sec:model-compare}

We first compare the fitting results of the alternative model functions for the BH mass term and the Eddington ratio term in \autoref{eq:rho:full}, and determine the fiducial model functions to use in our fitting of the joint distribution. For the Eddington ratio term, we test a Schechter function and a lognormal distribution function (\S\ref{parametricmodel}). The fitting results for the SDSS\_MO sample are shown in the left panel of \autoref{fig:model-compare} and listed in \autoref{tab:fitting-result} as models 1 and 2, respectively. In this test, we use the double power-law function for the mass term. In \autoref{fig:model-compare}, the black solid and blue dotted lines represent the best fitting models of the measured ERDF, using the Schechter and lognormal models as the Eddington ratio term, respectively. Both models produce largely consistent results with overlapping 1$\sigma$ regions.
The ERDF at large Eddington ratios ($\mathrm{log}(\lambda)\sim -0.5-0$) is well constrained by both models, and is consistent with the the $1/V_{\rm max}$ results. The uncertainties of the model constraints increase rapidly at both the high- and low-Eddington ratio end as a result from small sample statistics in these regimes. The limited sample statistics at the low-Eddington ratio end is mainly due to the flux limit in the selection function (also see left panel in \autoref{fig:bhmf-2d} on the effect of the flux limit on the ERDF). Nevertheless, the simple $1/V_{\rm max}$ method suffers the most from the flux limit, and under-predicts the abundance at the low Eddington ratio end.

The Schechter ERDF model has an asymmetric shape and is thus more flexible than the lognormal model. The Schechter model also results in a slightly smaller error range than the lognormal model. Therefore, we adopt the Schechter model for the Eddington ratio term in the following analysis.

To test the BH mass term, we compare a double power-law function and a modified Schechter function (\S\ref{parametricmodel}). The fitting results using the QLF fitting method for the luminosity sample are shown in the right panel of \autoref{fig:model-compare} and listed in \autoref{tab:fitting-result} as models 3 and  4, respectively. In this test, a Schechter model is used for the Eddington ratio term. The black solid and blue dotted lines represent the best fitting models of the measured QLF, using the double power-law model and the modified Schechter model as the mass term, respectively. The blue points and solid line represent the QLF results from  \citet{Matsuoka2018QLF}. The double power-law BHMF model results in a nearly identical QLF as the one in \citet{Matsuoka2018QLF}, while the modified Schechter model predicts a QLF with slightly steeper slopes at the low and high luminosity ends. Therefore, we adopt the double power-law model for the mass term in Eq.~\ref{eq:rho:full} in the following analysis. 

\subsection{2D Results from the Maximum Likelihood Method}
\label{sec:sdss-mo-results}

   \begin{figure*}
   \plottwo{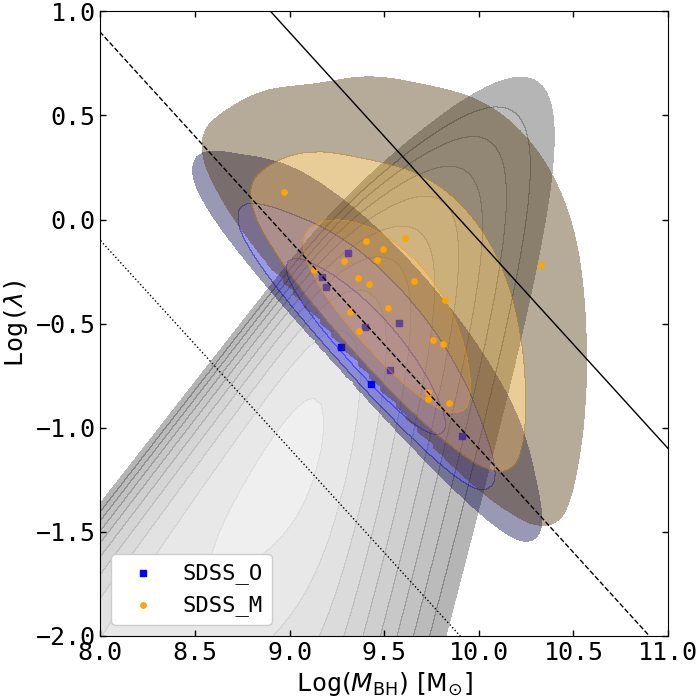}{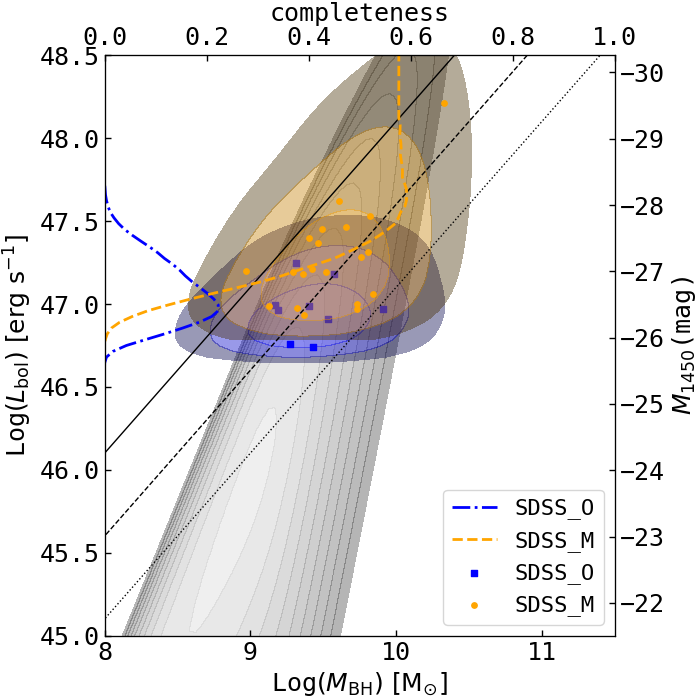}
   \caption{Intrinsic and observed bivariate distributions in the BH mass-Eddington ratio plane (left panel) and BH mass-luminosity plane (right panel).
   In both panel, the orange circle and blue squares represent individual quasars in the SDSS\_M and SDSS\_O fields, respectively. The shaded region with the same color demonstrates the observed (or expected) bivariate distribution for that field. The contour lines represent the $1\sigma$, $2\sigma$, and $3\sigma$ regions. The gray shaded region shows the shape of underlying intrinsic distribution from the best-fitting model. The gray contours show the constant space densities, from $10^{-12}$ to $10^{-7.5}$ in steps of $10^{0.5}$ ($\mathrm{Mpc}^{-3} ~ \mathrm{dex}^{-2}$). In the left panel, the dotted, dashed, and solid black lines show the constant bolometric luminosities of $10^{46}$, $10^{47}$, and $10^{48} \mathrm{erg~s}^{-1}$, respectively. In the right panel, the 
  dotted, dashed, and solid black lines show the Eddington ratios of $10^{-1}$, $10^{-0.5}$, and $10^{0}$, respectively. 
  We also show the luminosity selection functions (top axis) as function of luminosity (left or right axes) for the two samples (SDSS\_M and SDSS\_O) in orange dashed and blue dash-dot lines, respectively. Note these completeness functions do not have one-to-one correspondence to BH mass (i.e., the top and bottom axes are independent from each other). These are the original selection functions based on $M_{1450}$ in the corresponding quasar surveys \citep{Jiang2016SDSSQ}, marginalized over the $5.7<z<6.5$ redshift range. These (luminosity) selection function curves provide useful information on the luminosity range of observable quasars (i.e., the data points) in each survey, as well as the correction in quasar abundance due to the selection incompleteness in luminosity.
  }
   \label{fig:bhmf-2d}
   \end{figure*}
   
   
We perform the maximum likelihood fitting for the SDSS\_MO sample using the fiducial double power-law+Schechter model discussed in \autoref{sec:model-compare}. The  results are shown in \autoref{fig:bhmf-1d} and \autoref{fig:bhmf-2d} and listed as model 2 in \autoref{tab:fitting-result}. In \autoref{fig:bhmf-1d}, the intrinsic and measured distributions are plotted as the black solid and blue solid lines, respectively. The observed BHMFs of the SDSS\_M and SDSS\_O samples are represented by the orange dashed and blue dash-dot lines, respectively. Integrating our fiducial model over the survey volume, we obtain the predicted distributions of BH mass, Eddington ratio, and bolometric luminosity, as shown in \autoref{fig:bhmf-hist}.

In \autoref{fig:bhmf-hist}, the survey observed distribution (predicted sample distributions for one specific survey) of the SDSS\_O sample (blue dash-dot lines) as predicted by our model is lower than the observed data (blue histogram). This is because we normalized the model distribution using the sample size of SDSS\_M as reasoned in \autoref{sec:likelihood-method}. With this normalization, our best fitting model predicts 24 quasars in the SDSS\_M sample, which is the same number of the observed quasars in that sample. But the model only predicts 6.9 quasars in the SDSS\_O sample, less than the 10 objects observed in the SDSS\_O sample, albeit with small number statistics. If we use the full sample size to calculate the normalization factor, it will change by a factor of $(24+10)/(24+6.9) = 1.1$ or 0.04 dex, which is smaller than the $1\sigma$ error of the normalization from the QLF fitting method (\autoref{tab:fitting-result}, model 3). Since the discrepancy is very small, we stick to the normalization factor determined using the SDSS\_M sample alone.

In \autoref{fig:bhmf-2d}, we show the intrinsic and observed bivariate distributions in the BH mass-Eddington ratio plane (left panel) and BH mass-luminosity plane (right panel) for our best-fit model. They are calculated by marginalizing over redshift for the observed distribution (\autoref{eq:psi:obs}). In both panels, the orange and blue points represent the measurements of individual quasars in the SDSS\_M and SDSS\_O samples. Most of the data points are located in the $1\sigma$ region of their respective sample and all of them are within the $3\sigma$ region, indicating good agreement with our best-fitting model. The intrinsic distribution is displayed as the gray shaded region, which is modified to the observed distribution after convolving with the error model and incorporating the survey selection function.
 
In both panels of \autoref{fig:bhmf-2d}, there is a boundary in the observed distributions towards the low-luminosity end, corresponding to the flux limit of each sample. In the mass-luminosity plane, we also plot the redshift-marginalized selection functions of the SDSS\_M and SDSS\_O samples as the orange dashed and blue dash-dot lines, to demonstrate the decline of the selection probability near the flux limit.

	\begin{landscape}
	\begin{table}
		\begin{threeparttable}
			\caption{Models and their fitting results}
			\label{tab:fitting-result}
			\begin{tabular}{ccccccccccccc}
				\toprule
				\multicolumn{3}{c}{} & \multicolumn{4}{c}{BH mass term} & \multicolumn{4}{c}{Eddington ratio term} & \multicolumn{2}{c}{Error model}\\
				\cmidrule(l){4-7} \cmidrule(l){8-11} \cmidrule(l){12-13}
				{ID} & {sample}  & {\makecell[ct]{model$^a$ \\ (BH mass+ER term)}}  & 
				{$\log \Psi_*$} & {$\log M^*_\bullet$} & {$\alpha$}  & {$\beta$} & 
				{$k_\lambda$} & {$\log \lambda_0$}  & {$\sigma_\lambda$} & {$\alpha_\lambda$} & 
				{$\beta_e$} & {$\sigma_{\mathrm{VM}}$} \\
				\midrule
				1&SDSS\_MO&DPL+lognormal&$-4.97^{+3.70}_{-2.81}$&$8.43^{+0.89}_{-0.99}$&$-1.12^{+0.77}_{-0.74}$&$-4.40^{+0.96}_{-1.02}$&$0.98^{+0.77}_{-0.51}$&$-0.63^{+0.13}_{-0.18}$&$0.23^{+0.13}_{-0.12}$&-&0.20&0.25\\
				2&SDSS\_MO&DPL+SCH&$-4.41^{+3.76}_{-2.71}$&$8.47^{+0.76}_{-0.95}$&$-1.09^{+0.74}_{-0.75}$&$-4.82^{+0.96}_{-0.81}$&$1.35^{+0.71}_{-0.59}$&$-0.81^{+0.29}_{-0.19}$&-&$0.77^{+0.79}_{-0.86}$&0.20&0.25\\
				3&luminosity&DPL+SCH&$-6.47^{+0.15}_{-0.21}$&$9.04^{+0.11}_{-0.10}$&$-0.93^{+0.50}_{-0.57}$&$-4.47^{+0.42}_{-0.51}$&1.13&-0.85&-&0.71&0.20&0.25\\
				4&luminosity&MSCH+SCH&$-6.23^{+0.31}_{-0.40}$&$8.89^{+0.25}_{-0.32}$&$-1.42^{+0.61}_{-0.54}$&$0.89^{+0.21}_{-0.17}$&1.13&-0.85&-&0.71&0.20&0.25\\
				5&SDSS\_MO&DPL+SCH&$-5.97^{+3.71}_{-1.67}$&$8.72^{+0.64}_{-1.03}$&$-1.10^{+0.73}_{-0.73}$&$-4.33^{+0.92}_{-1.04}$&$0.95^{+0.70}_{-0.51}$&$-0.83^{+0.33}_{-0.34}$&-&$0.79^{+0.71}_{-0.67}$&$0.20^{+0.48}_{-0.41}$&$0.22^{+0.06}_{-0.06}$\\
				6&SDSS\_MO&DPL+SCH&$-4.71^{+3.95}_{-2.52}$&$8.62^{+0.70}_{-1.00}$&$-1.08^{+0.74}_{-0.78}$&$-4.81^{+0.95}_{-0.80}$&$1.36^{+0.77}_{-0.60}$&$-0.93^{+0.23}_{-0.18}$&-&$0.68^{+0.81}_{-0.96}$&0.00&0.25\\
				\bottomrule
			\end{tabular}
			\begin{tablenotes}
				\small\item Fitting results of all models. A parameter is fixed when it has no errors. A parameter value and its $1\sigma$ region are the 50th, 16th and 84th of the posterior distribution. Model 2 is the fiducial model of the BH mass sample and Model 3 is the fiducial model of the QLF sample. Model 3 uses the same Eddington ratio term as the best-fit Model 2, but the best-fit parameters are different with the 50th posterior distribution. For example, for model 2, the 50th posterior distribution value of $k_\lambda$ is $1.35$, while the besting-fit model gives $1.13$.\item $^a$ DPL is the double power-law model. SCH is the Schechter model. MSCH is the modified Schechter model.\end{tablenotes}
		\end{threeparttable}
	\end{table}
	\end{landscape}

The BHMF fitting results are shown in the left panel of \autoref{fig:bhmf-1d}. Comparing the intrinsic and measured distributions, the virial BH mass uncertainties (determined by $\sigma_{\mathrm{VM}}$ and $\beta_e$) result in the overestimation of the BHMF at the high-mass end in the measured distribution. Using the SDSS\_MO sample, we derive reasonably tight constraints on the intrinsic BHMF in the $M_{\rm BH} \gtrsim 10^{9.3}~M_\odot$ regime. The intrinsic BHMF below $10^{9.3}~M_\odot$ is poorly constrained due to the limited number of low-luminosity (and low-mass) quasars in our sample. We will improve the BHMF constraints using the QLF fitting method and the larger luminosity sample in \autoref{sec:lf-results}.

The BHMF calculated by the $1/V_{\rm max}$ method is shown as the orange circle and blue squares in \autoref{fig:bhmf-1d} and listed in \autoref{tab:vmax-bhmf} in Appendix. In the region of $M_{\rm BH}>10^{9.5}~M_\odot$, they overlap with the $1\sigma$ region of the measured model BHMF. In the low-mass region, however, the $1/V_{\rm max}$ BHMF suffers from the selection incompleteness and the turnover is artificial. Here we emphasize that the maximum likelihood results are not a fit to the $1/V_{\rm max}$ binned BHMF. The binned $1/V_{\rm max}$ BHMF corrects for some selection incompleteness due to the flux limit, which is why it lies above the model-predicted observed BHMF (orange dashed and blue dash-dot lines). Since the $1/V_{\rm max}$ method does not properly correct for incompleteness in terms of BH mass, the binned BHMF is only shown for comparison and not recommended for further use.

The QLF model predictions are shown in the right panel of \autoref{fig:bhmf-1d}. The minor differences between the intrinsic and measured distributions are caused by the luminosity scatter ($\sigma_{\mathrm{bol}}$). There is good agreement between our results and the QLF in  \citet{Matsuoka2018QLF}. Again, the QLF below $-26$ mag is poorly constrained, which is due to the limited luminosity range of our BH mass sample. The $1/V_{\rm max}$ binned QLF data (listed in \autoref{tab:vmax-lf} in Appendix) are also shown for reference in orange circle and blue squares in the figure.
 
The ERDF model predictions are shown in the middle panel of \autoref{fig:bhmf-1d}. The differences between the intrinsic and measured distributions are caused by the luminosity and virial BH mass uncertainties. The $1/V_{\rm max}$ binned ERDF shown in orange circle and blue squares is highly incomplete at $\log \lambda < -0.5$. The best-fitting model also shows blown-up uncertainties at the low Eddington ratio end. The best-fit model ERDF will be used in our QLF-fitting application to the larger luminosity sample.

\begin{figure}
   \plotone{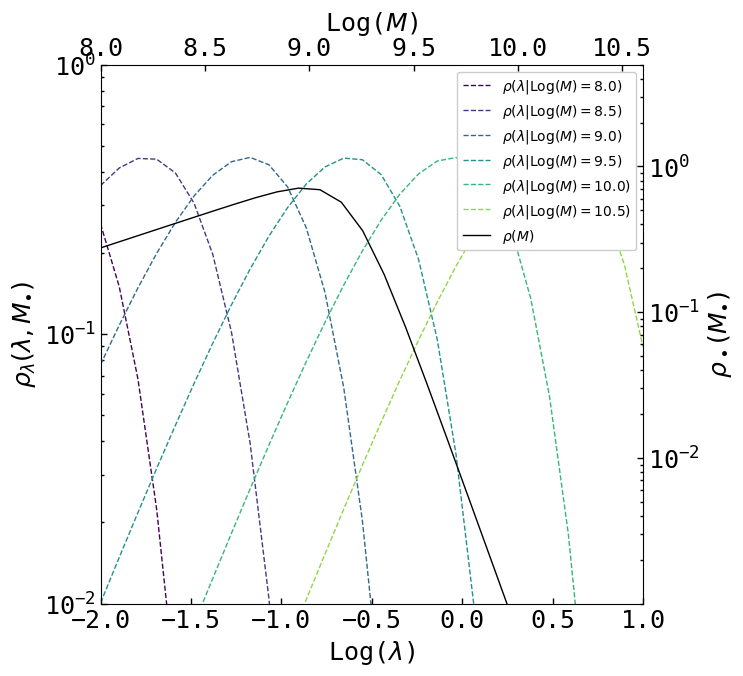}
   \caption{
    The BH mass term and Eddington ratio term of the best-fit model. The black line shows the double power-law BH mass term. The color-coded lines show the mass dependent Schechter Eddington ratio term. The different colors show the conditional ERDF at different masses from $10^8$ to $10^{10.5} \mathrm{M}_\odot$.
   }
   \label{fig:erdf-term}
   \end{figure}

In \autoref{fig:erdf-term}, we show the conditional ERDF at different BH masses for our best-fit model ($\rho_\lambda(\lambda, M_\bullet)$, \autoref{eq:rho:lambda}). Similar to the lognormal model, the Schechter profile with $\alpha_{\lambda}>-1$ features a broad peak around a characteristic value (our best-fit model has a positive $\alpha_{\lambda}$; see Table \ref{tab:fitting-result}). We have a positive $k_\lambda \sim 1.1$ in our best-fit model (see \autoref{tab:fitting-result}) and thus the break point shifts to lower Eddington ratios for lower masses. The 1D ERDF (\autoref{fig:bhmf-1d}, middle panel) is calculated by integrating over $M_\bullet$ in the $\lambda - M_\bullet$ plane. The fiducial integration range in mass is $\log M_\bullet=8.5-10.5$, for which the peak Eddington ratio ranges from $\log\lambda\approx -1.7$ to $\approx 0.5$ (\autoref{fig:erdf-term}). Therefore, the resultant intrinsic 1D ERDF displays a broad peak around $\log\lambda\approx -0.9$ (middle panel of \autoref{fig:bhmf-1d}). For comparison, \citet{Schulze2015BHMF} obtained $k_\lambda \approx 0.10-0.15$ for a much larger sample at low redshift. Compared with their samples, our SDSS samples have smaller ranges in BH mass and Eddington ratio. The large value of $k_\lambda$ in our best-fit model is determined by objects with BH masses around $\log M_{\bullet,0}=9.5$ (\autoref{eq:lambda-star}) and should only hold for our sample with a small Eddington ratio range. A larger sample is needed to further test the BH mass dependence of Eddington ratios over a broader parameter space.

\subsection{Results from the Luminosity Sample}
\label{sec:lf-results}

\begin{figure}
   \plotone{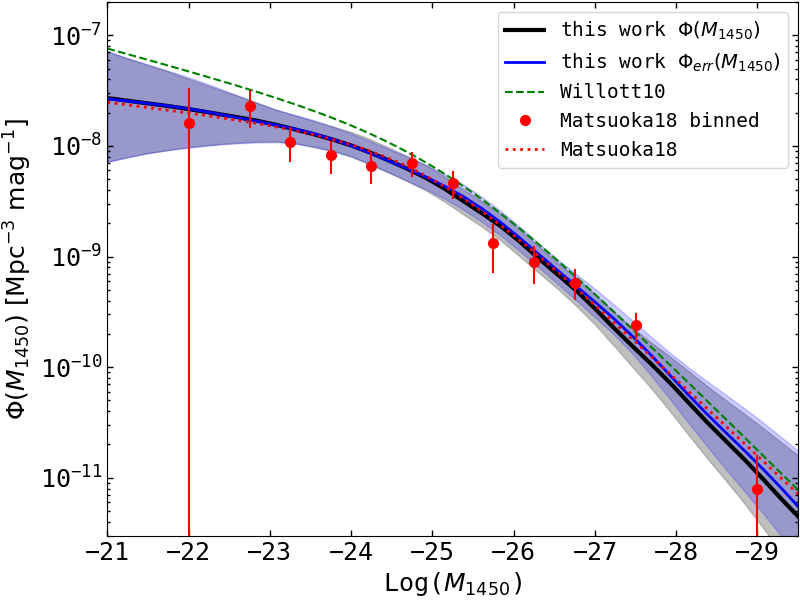}
   \caption{LF results of the QLF fitting method for the luminosity sample. All symbol are the same as the right panel in \autoref{fig:bhmf-1d}. The results (black solid line) are the same as those in \citet{Matsuoka2018QLF} (red dotted line).
   }
   \label{fig:lm-lf}
   \end{figure}

\begin{figure}
   \plotone{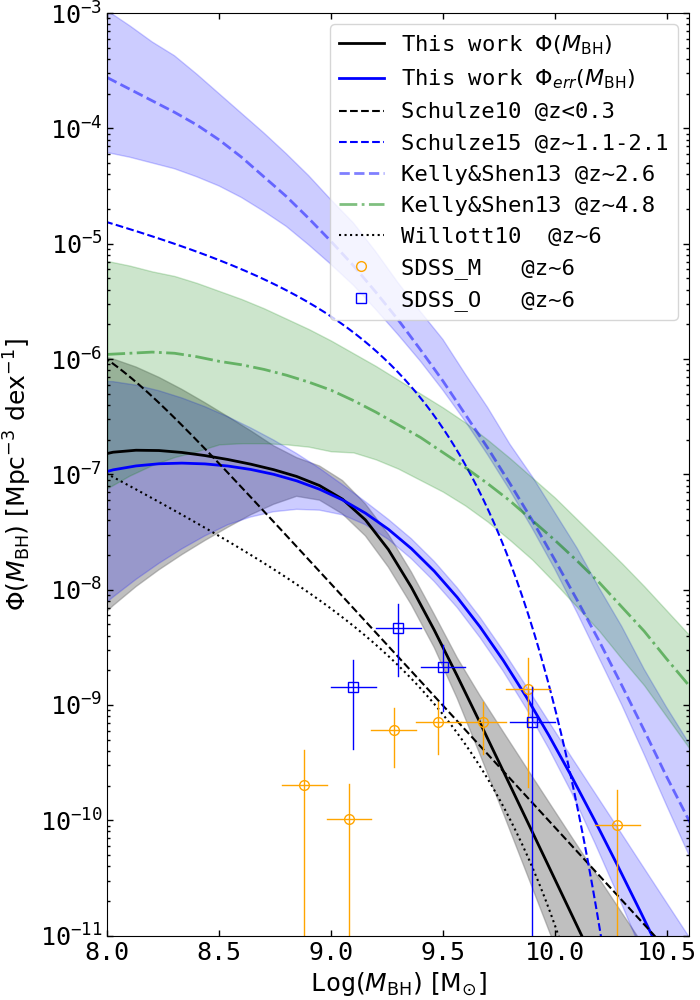}
   \caption{BHMF results of the QLF fitting method for the luminosity sample. The $1/V_{\rm max}$ BHMF, and the intrinsic and measured BHMF distributions with their $1\sigma$ regions have the same symbols as shown in the left panel in \autoref{fig:bhmf-1d}. We also collect active BHMF functions from previous studies. They are from \citet{Schulze2010BHMF}, \citet{Willott2010BHMF}, \citet{KellyShen2013BHMF} and \citet{Schulze2015BHMF}.
   }
   \label{fig:lm-bhmf}
   \end{figure}

Now we proceed to use the QLF-fitting method to improve the constraints on the BHMF based on the larger luminosity sample. We use the double power-law model for the intrinsic BH mass term and the same Eddington ratio term in the best-fitting model as in \autoref{sec:sdss-mo-results} (model 2 in \autoref{tab:fitting-result}). The fitting results are displayed in \autoref{fig:lm-lf} and \autoref{fig:lm-bhmf}, and listed as model 3 in \autoref{tab:fitting-result}. Our model is still the bivariate distribution (\autoref{eq:rho:full}) with a fixed Eddington ratio term, which is different from \citet{Willott2010BHMF} who applied a BHMF model and a fixed ERDF model.

\autoref{fig:lm-lf} shows that we successfully reproduced the QLF in \citet{Matsuoka2018QLF}, with the model QLF constructed from the underlying BH mass and Eddington ratio distributions. \autoref{fig:lm-bhmf} shows the final intrinsic BHMF that we derive for $z\sim 6$ quasars using the luminosity sample, along with results in previous studies. At the high-mass end, the results are similar to that from the maximum likelihood fitting to the BH mass sample (left panel of \autoref{fig:bhmf-1d}). In the low-mass regime, the QLF-fitting model is better constrained to $M_{\rm BH} \sim 10^{8.5}M_\odot$ than the previous results based on the BH mass sample. The usable region of the BHMF is thus larger than that in \autoref{fig:bhmf-1d}. The model data of BHMF and QLF are also listed in \autoref{tab:bhmf-3} and \autoref{tab:lf-3} in Appendix. Model 3 represents our final fiducial results for the 1D intrinsic BHMF and QLF.

\subsection{The Mass--Luminosity Plane of $z\sim 6$ Quasars}
The abundance of quasars in the 2D mass--luminosity plane is shown in \autoref{fig:bhmf-2d} (right). Similar to earlier studies in \citet{ShenKelly2012BHMF} and \citet{KellyShen2013BHMF}, our forward modeling is able to reveal the population of quasars below the flux limit of the survey. In addition, by accounting for uncertainties in the measured physical quantities (i.e., BH mass and luminosity), we are able to constrain the intrinsic distribution of quasars in the mass-luminosity plane. 
For example, we can constrain the abundance of $>10^{10}\,M_\odot$ BHs in $z\sim 6$ quasars to be
$0.53^{+1.27}_{-0.45} \times 10^{-11} \mathrm{Mpc}^{-3}$, or $0.52^{+1.25}_{-0.44}$ quasars in the 11,240 $\mathrm{deg}^2$ SDSS main survey, by integrating the BHMF. This result suggests there are a few $z\sim 6$ quasars with $M_{\rm BH}>10^{10}\,M_\odot$ over the full sky.

\section{Discussion}
\label{sec:discussion}

\subsection{Comparison with Previous Studies at $z\sim 6$}
\label{sec:result-compare}

 \citet{Willott2010BHMF} (hereafter W10) estimated the BHMF of $z\sim 6$ quasars using the QLF fitting method and an assumed Eddington ratio distribution.
 Their luminosity sample contains 40 quasars at $5.74<z<6.42$, selected from the known quasars in the SDSS main, S82 region, and the CFHQS field at that time. They used the Schechter and lognormal distributions to model the BHMF and ERDF, respectively. They also assumed that the intrinsic ERDF and the survey observed ERDF are both lognormal distributions. The offset of the peak Eddington ratio between the intrinsic and survey observed distributions was determined by simulations. Their active BHMF is shown as the black dotted line in \autoref{fig:lm-bhmf}.
 
Our method is different from the W10 method in several aspects.
First, we have an extra error model in the fitting process. It includes the scatters of virial masses and bolometric luminosity uncertainties. 
Our model QLF is generated from the bivariate distribution in the BH mass-Eddington ratio plane. 
In W10, the model QLF was the convolution of the model BHMF and ERDF, without extra error models.
In our QLF fitting method, the Eddington ratios are derived from the BH masses. Therefore the errors of the BH masses are propagated into the ERDF, requiring an error model. The measured and intrinsic distributions (blue and black solid lines in \autoref{fig:lm-bhmf}) demonstrate the differences with and without the error model. The uncertainty in virial BH masses leads to the overestimation of the BHMF at the high-mass end. The model measured QLF can also be generated by convolving our measured BHMF and ERDF. In this case, our measured BHMF result is equivalent to the BHMF of W10.

Second, our BHMF is higher in the $M_\bullet>10^{8.0} M_\odot$ region than the W10 results, which originates from the different Eddington ratio distribution models that we used. In W10, the intrinsic Eddington ratio distribution model had a lognormal distribution centered at $\log \lambda=-0.22$ (the break point).
In our model, we used an Eddington ratio distribution in the form of a Schechter function with a break point $\log \lambda_*=-0.85$ over $\log M_\bullet=8.5-10.5$. Qualitatively, a lower Eddington ratio break point results in a higher BH mass break ($M^*_\bullet$), which will shift the BHMF to higher masses (to the right). In \autoref{fig:lm-bhmf}, when we shift the measured BHMF (blue line) by $\sim-0.5$ dex, we obtain a very similar shape with the W10 BHMF (the black dotted line). The final QLF models of W10 and ours are close to each other, especially in the high-luminosity end (\autoref{fig:lm-lf}, green dashed and black solid lines). Therefore, the main difference between our BHMF and that in W10 is caused by the difference in the ERDF break point. 

\subsection{Choices of Virial BH Mass Estimates}

\begin{figure}
   \includegraphics[width=0.5\textwidth]{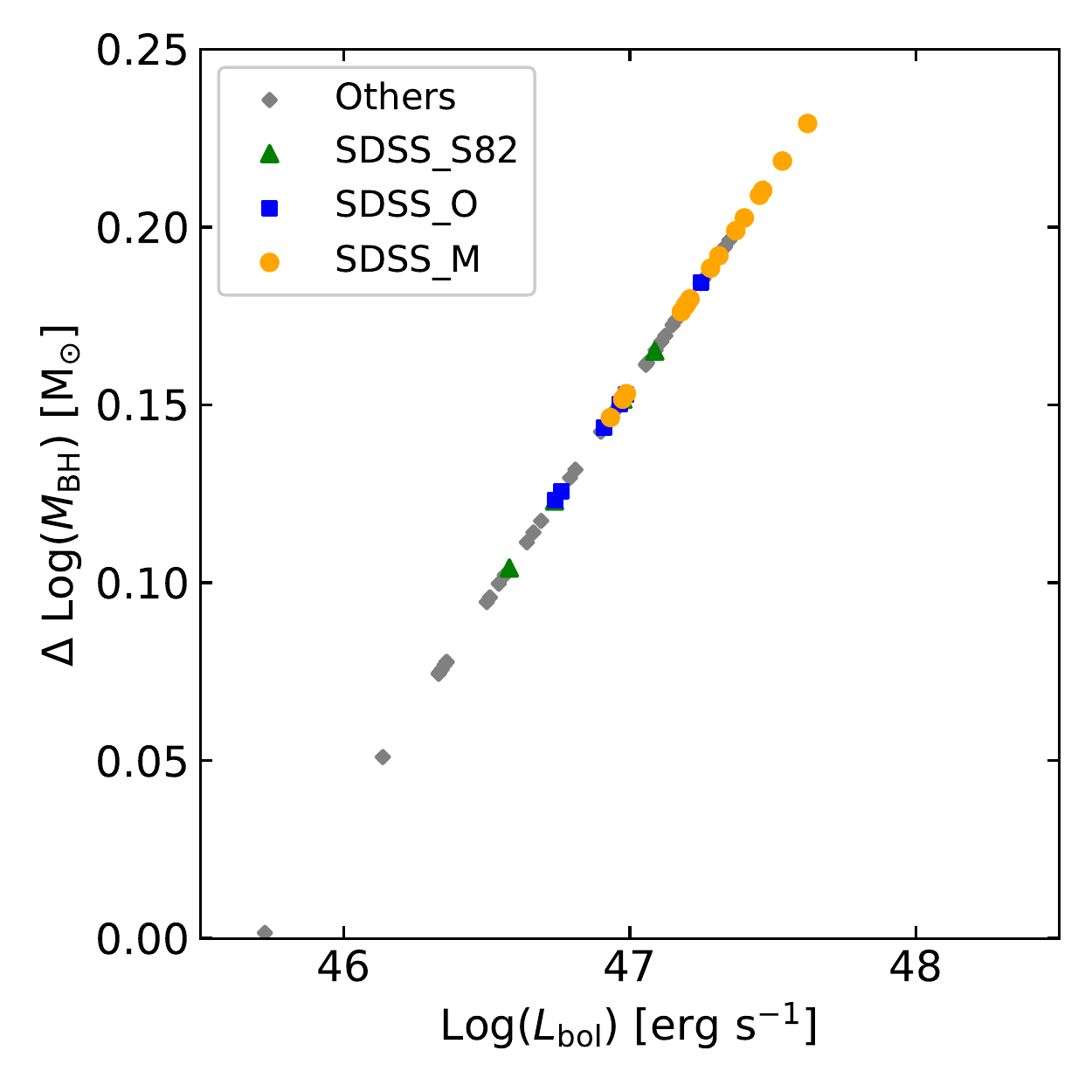}
   \caption{Differences in Mg II-based BH masses estimated from the recipes of \citet{Shen2011C} (our fiducial recipe) and \citet{Vestergaard_Osmer_2009}. On average, our BH masses are $\sim 0.2$ dex higher than those estimated using the \citet{Vestergaard_Osmer_2009} recipe.
   The symbols have the same meanings as in \autoref{fig:sample-2d}.
   }
   \label{fig:recipe-diff}
   \end{figure}

In \citet{Shen2019GNIRS}, the \MgII-based virial BH mass recipe adopted was calibrated to match the \hbeta- and \CIV-based recipes of \citet{Vestergaard_Peterson_2006}  using SDSS quasar samples \citep{Shen2011C}. This is also the fiducial recipe that we used for our BH mass sample. In our study, we collected near-IR spectral fitting results from the literature. \citet{Jiang07BHM}, \citet{DeRosa2011BHM}, and \citet{Wu15QSO} utilized the \MgII\ recipe of \citet{McLure_Dunlop_2004} which produces lower BH masses by 0.22 dex on average than our fiducial \MgII-based recipe. W10 applied the \MgII-based recipe of \citet{Vestergaard_Osmer_2009}, which has a shallower luminosity dependence than our recipe does, and would yield smaller masses (higher Eddington ratios) on average for luminous $z\sim 6$ quasars. In \autoref{fig:recipe-diff}, we plot the differences in BH masses estimated using the \citet{Shen2011C} and \citet{Vestergaard_Osmer_2009} recipes, for quasars with \MgII\ FWHM measurements. Compared to \citet{Vestergaard_Osmer_2009}, the recipe of \citet{Shen2011C} produces $\sim 0.2$ dex higher masses for our SDSS sample. These differences in BH mass estimates contribute to the ERDF differences between our work and W10, but the main difference in the ERDF should come from the different fitting methods.

\subsection{Comparison with Previous Methods}

As discussed in \S\ref{sec:method}, there are two methods (both based on the likelihood analysis) used to fit the BHMF. Our likelihood method fits the data in the BH mass--Eddington ratio plane, while the Bayesian framework of \citet{KellyShen2013BHMF} performed their fitting in the mass--luminosity plane. Since the Eddington ratio is derived from mass and luminosity, these two procedures are equivalent and both methods can produce the observed bivariate distribution in the mass-luminosity plane (\autoref{fig:bhmf-2d}, right panel). Our error model is an updated version of that in \citet[]{Schulze2015BHMF}. We added a $\beta_e$ parameters to account for the luminosity-dependent bias in single-epoch virial masses. \citet{KellyShen2013BHMF} also have this parameter in their models. One key difference between our method and \citet{KellyShen2013BHMF} is the parametric functions for the intrinsic distributions. In our fiducial model, we used a double power-law and a Schechter function to model the mass term and the Eddington ratio term, respectively. \citet{KellyShen2013BHMF} used a mixture of five 2D lognormal distributions to model the bivariate distribution. The mixture is flexible enough to capture the basic shape of any physical BHMF and largely simplifies the computation as many integrations can be done analytically, but the number of parameters is much larger than that in our modeling. For our small sample size, simplified models with fewer parameters are sufficient to describe the data and to avoid overfitting problems.

\subsection{Luminosity-dependent Bias in Virial BH Masses}
\label{sec:errors}

In our fiducial model described in \S\ref{sec:likelihood-method}, we adopted the error model (\autoref{eq:error-model}) with a fixed $\sigma_{\mathrm{VM}}=0.25$ and $\beta_e=0.2$. If we fit these two parameters as free parameters in the model, the results are listed as model 5 in \autoref{tab:fitting-result}.
The best-fit parameters are $\sigma_{\mathrm{VM}}=0.23^{+0.05}_{-0.07}$ and $\beta_e=0.21^{+0.47}_{-0.42}$.
While these constraints are not tight (particularly for $\beta_e$), as expected due to the small sample size, these best-fit parameters are reasonable and consistent with the constraints in \citet{ShenKelly2012BHMF}. For completeness, we also test a model assuming $\beta_e=0$, and list the results as model 6 in \autoref{tab:fitting-result}. There are no significant differences in the final BHMFs between models 5/6 and model 2, and all properties of the model predictions remain qualitatively unchanged.

   \begin{figure}
   \includegraphics[width=0.5\textwidth]{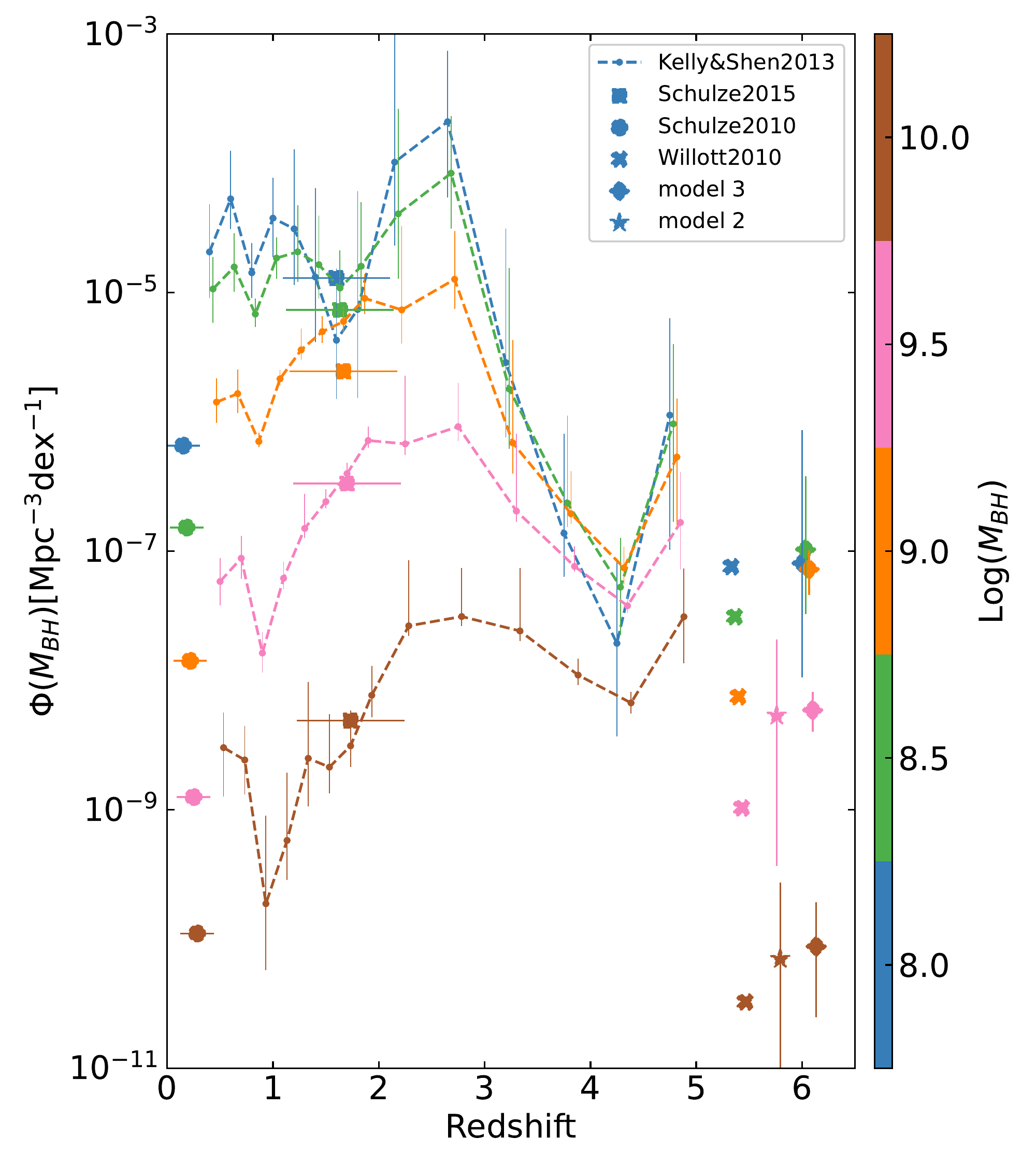}
   \caption{Redshift evolution of the quasar number density at different BH masses. Different colors represent different mass ranges. 
   Results from \citet{Schulze2010BHMF} and \citet{Schulze2015BHMF} are shown in circles and squares, respectively. The error bars in the horizontal direction show the redshift coverages of their samples. The \citet{KellyShen2013BHMF} results are displayed as the dashed lines with point markers. The error bars are calculated according to their 68\% percentile range of the BHMFs. For clarity, many data points have been slightly shifted horizontally.
   }
   \label{fig:z-evolution}
   \end{figure}

\subsection{Evolution of Quasar Abundance}
An important result in quasar demographics is the cosmic downsizing evolution, i.e., the number density of less luminous objects peaks at lower redshift. It was initially discovered in the X-ray surveys \citep[e.g.,][]{Cowie2003, Hasinger_etal_2005} and then confirmed in optical observations \citep[e.g.,][]{Croom_etal_2009, ShenKelly2012BHMF}.
Recently, \citet{ShenXuejian2020} measured bolometric QLFs at $z = 0-7$. They found that the bolometric QLF rises with time monotonically at $z \gtrsim 2-3$, following the hierarchical structure formation paradigm \citep[e.g.,][]{Shankar_etal_2009}. At $z \lesssim 2-3$, the QLF stops rising with time and shows a continuous horizontal shift towards the low luminosity regime. They also reported flatter bright-end LF slopes at $z \gtrsim 2-3$. For quasars at higher redshift ($z\gtrsim4$), recent QLF studies found no strong redshift evolution for the bright-end slope $\beta$ \citep[e.g.,][]{Akiyama2018,Matsuoka2018QLF,Jiang2016SDSSQ, Kim2020QLF, WangFeige2019b}.  

Besides the downsizing evolution of the QLF, there is evidence that BHMF also has a downsizing trend \citep[e.g.,][]{Vestergaard_Osmer_2009, ShenKelly2012BHMF, KellyShen2013BHMF}. In \autoref{fig:z-evolution}, we plot the quasar number densities in different mass bins as a function of redshift from several BHMF studies. \citet[]{KellyShen2013BHMF} studied BHMFs of SDSS quasars at $z= 0.4 \sim 4.75$, and their results are displayed as the dashed lines. They found that the peaks of the number densities are around $z=2\sim3$. At $1 \lesssim z \lesssim 2.5$, the number densities of more massive BHs fall off more rapidly with decreasing redshift. At $2.5 \lesssim z \lesssim 4.5$, the number densities of more massive BHs fall off more slowly with increasing redshift. 

We also include the results of our $z\sim6$ BHMF in \autoref{fig:z-evolution}. We find that the evolutionary trend at $4.5 \lesssim z \lesssim 6$ is contrary to the trend at $2.5 \lesssim z \lesssim 4.5$. At $4.5 \lesssim z \lesssim 6$, the number densities of more massive BHs decline more rapidly with increasing redshift. As the uncertainties of the abundance measurements are still large, future samples with more BH mass measurements are needed to confirm this result.

\section{Summary}
\label{sec:summary}

We have presented the measurement of the demographics of $z\sim 6$ quasars in the BH mass-luminosity plane using the largest available sample of quasars with well defined selection functions in this redshift regime. With a forward modeling approach, we were able to constrain the intrinsic distribution of quasars by accounting for the selection completeness of the observed sample and the uncertainties in the measured BH masses and luminosities. In particular, we provided robust constraints on the SMBH abundance at $M_\bullet\gtrsim 10^{8.5}\,M_\odot$. The Eddington ratios of these $M_\bullet\gtrsim 10^{8.5}\,M_\odot$ quasars can be approximated by a mass-dependent Schechter model with a broad peak around $\log\lambda\sim -0.9$.
With our model intrinsic distributions, we constrain the abundance of $>10^{10}\,M_\odot$ quasars to be $0.53^{+1.27}_{-0.45} \times 10^{-11} \mathrm{Mpc}^{-3}$ at $z\sim 6$. This predicts a total of $0.52^{+1.25}_{-0.44}$ quasars (above $10^{10}\,M_\odot$) in the 11,240 $\mathrm{deg}^2$ SDSS main survey of $5.7<z<6.5$ quasars.

Comparing the BHMF results with previous studies, we find that the evolution of $M_{\mathrm{BH}}\gtrsim10^{9.5} M_{\odot}$ quasars is faster than that of $10^{8} \lesssim  M_{\mathrm{BH}} \lesssim10^{9.5} M_{\odot}$ quasars at $z\sim 6$. The abundance of the most massive active SMBHs ($M_{\mathrm{BH}} \gtrsim10^{9} M_{\odot}$) is much lower at $z\sim6$ than their counterparts at lower redshifts, reflecting the early build up of this population.

For future work, we will apply this methodology to deeper samples of $z\gtrsim 6$ quasars with improved sample statistics and BH mass measurements, such as quasar samples from the Chinese Space Station Telescope slitless spectroscopic survey \citep{Zhan_2021}. This will allow us to probe the more common population of quasars/SMBHs with lower luminosities and lower BH masses at cosmic dawn. Improved constraints of BHMF and ERDF from these deeper quasar samples will be necessary to study the population of $z\gtrsim 6$ quasars in the low-luminosity and low-mass regime, and shed light on the assembly of these earliest SMBHs, such as the origin of BH seeds and the evolution of accretion rate during BH growth.

\section*{Acknowledgements}
\addcontentsline{toc}{section}{Acknowledgements}

We thank the anonymous referee for useful comments that improved the manuscript, C. Willott and Y. Matsuoka for providing the selection functions in the CFHQS and SHELLQs surveys, and W. He for helpful discussions. We acknowledge support from the National Key R\&D Program of China (2016YFA0400703), the National Science Foundation of China (11721303, 11890693), and the science research grants from the China Manned Space Project with NO. CMS-CSST-2021-A05. Y.S. acknowledges support from NSF grants AST-1715579 and AST-2009947.
M.V. gratefully acknowledges financial support from the Independent Research Fund Denmark via grant number DFF 8021-00130.
F.W. thanks the support provided by NASA through the NASA Hubble Fellowship grant \#HF2-51448 awarded by the Space Telescope Science Institute, which is operated by the Association of Universities for Research in Astronomy, Incorporated, under NASA contract NAS5-26555. 

\section*{DATA AVAILABILITY}
All catalogue data used in this paper are publicly available and the details are presented in \autoref{tab:all-samples} and \autoref{tab:bhm-sample}.

\bibliographystyle{mnras}
\bibliography{BHMF,refs}

\clearpage
\appendix

\counterwithin{figure}{section}
\counterwithin{table}{section}

\section{Integration ranges for the BHMF and ERDF}
\label{app:a}
\begin{figure*}
   
   \plottwo{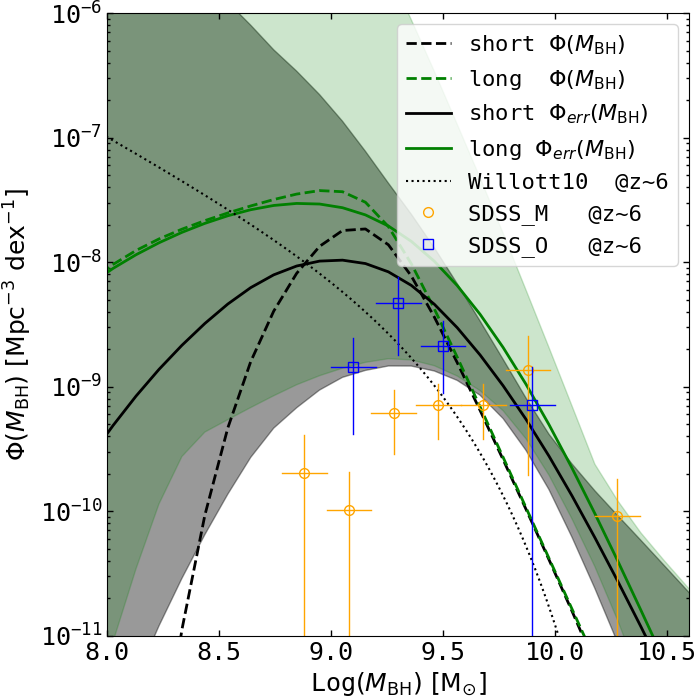}{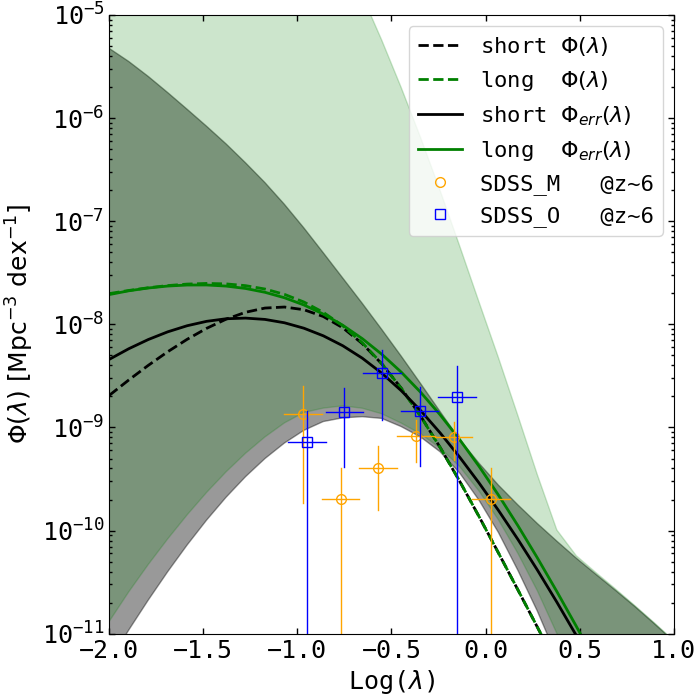}
   \caption{The black and green solid lines show the measured distributions calculated over the short and long integration ranges, respectively. The shadow regions with the same colors show their $1\sigma$ regions. The black and green dashed lines show the corresponding intrinsic distributions. Other symbols are the same as shown in the left panel of \autoref{fig:bhmf-1d}. The long and short ranges mean the broad integration ranges ($-2.5<\log \lambda<0.5$, $6.0<\log M_\bullet<10.5$) and narrow integration ranges ($-1.2<\log \lambda <0.5$, $8.5<\log M_\bullet<10.5$) used to calculate the 1D distributions.
   }
   \label{fig:int-range}
   \end{figure*}

For the maximum likelihood method, the 2D BH mass-luminosity plane and mass-Eddington ratio plane provide rich demographic information about $z\sim 6$ quasars, while the 1D BHMF, ERDF, and QLF distributions are marginalized distributions of the 2D distributions. The shapes of these 1D functions may depend on their integration ranges over other variables. To investigate this effect, we compare the BHMF and ERDF calculated using different integration ranges in \autoref{fig:int-range}.
These 1D functions are calculated using either narrow integration ranges ($-1.2<\log \lambda <0.5$, $8.5<\log M_\bullet<10.5$) or broad integration ranges ($-2.5<\log \lambda<0.5$, $6.0<\log M_\bullet<10.5$). The narrow range only covers the range of our observed data while the broad range covers a more extended parameter space for the integration to converge. The black and green solid lines show the measured distribution calculated over the narrow and broad integration ranges, respectively. The shaded regions with the same colors show the $1\sigma$ uncertainty regions. The black and green dashed lines show the corresponding intrinsic distributions. Other symbols are the same as shown in \autoref{fig:bhmf-1d} (left panel).

Comparing the measured distributions over the narrow and broad integration ranges, we find that narrower integration ranges result in lower 1D distribution functions, especially at the low-value end. However, the effects on the BHMF and ERDF are different. As shown in \autoref{fig:erdf-term}, the Schechter Eddington ratio term increases with $\lambda$ below the break point. The total probability is converged for a Schechter function with $\alpha_\lambda>-1$ and the BHMF (calculated by integrating the bivariate distribution over $\lambda$) is also converged when the integration range (over $\lambda$) is large enough. In our fitting result, the $1\sigma$ region of $\alpha_\lambda$ is $>-1$ and the integral is converged for BHMF. On the other hand, the double power-law BH mass term decreases monotonically with mass when $\alpha<-1$, and the ERDF (calculated by integrating the bivariate distribution over $M_{\bullet}$) continues to increase when lower and lower BH masses are included in the integration. In our fitting result, nearly half of the models have $\alpha<-1$ within 1$\sigma$, which results in increasing ERDF when lower BH masses are included in the integration range.

In the main text, the integration range of $\lambda$ is large enough to ensure that our BHMF is roughly converged. For the ERDF, we adopt a limited integration range of $8.5<\log M_\bullet<10.5$, and the ERDF should be considered as the Eddington ratio distribution function corresponding to this particular BH mass range (roughly consistent with the observed BH mass range in our sample). 

\section{Data tables}
\label{app:b}

	\begin{table*}
		\begin{threeparttable}
			\caption{Quasar samples used in this work}
			\label{tab:all-samples}
			\begin{tabular}{cccc}
				\toprule
				{Sample} & {subsample} & {Luminosity Sample} & {BH mass sample} \\
				\midrule
				SDSS & SDSS\_M & 24 & 20 \\
				SDSS & SDSS\_O & 10 & 9 \\
				SDSS & SDSS\_S82 & 13 &  \\
				CFHQS & - & 17 &  \\
				SHELLQs & - & 48 &  \\
				total & - & 110$^a$ & 29 \\
				\bottomrule
			\end{tabular}
			\begin{tablenotes}
				\small
				\item The Luminosity Sample is from \citet{Matsuoka2018QLF}. See \autoref{tab:bhm-sample} for details of the BH mass sample.
				\item $^a$ {Quasar J231546.58-002357.9 exists in all three surveys.}
			\end{tablenotes}
		\end{threeparttable}
	\end{table*}

	\begin{table*}
		\begin{threeparttable}
			\caption{The BH mass sample}
			\label{tab:bhm-sample}
			\begin{tabular}{lllccccc}
				\toprule
				{OBJID} & {survey} & {BH Ref$^a$} & {Redshift} &{Source}& {$\mathrm{log}(L_{\mathrm{bol}})$} & {$\mathrm{log}(M_{\mathrm{BH}})$} & {$\mathrm{log}(\lambda)$} \\
				\midrule
				J0002+2550 & SDSS\_M & Wang21 & 5.82 & \MgII & $47.180\pm0.003$ & $9.36\pm0.05$ & $-0.28\pm0.05$\\
				J0100+2802 & SDSS\_M & Wu15 & 6.30 & \MgII & $48.210\pm0.065$ & $10.33\pm0.07$ & $-0.22\pm0.09$\\
				J0810+5105 & SDSS\_M & Wang21 & 5.81 & \MgII & $47.193\pm0.009$ & $9.29\pm0.11$ & $-0.20\pm0.11$\\
				J0836+0054 & SDSS\_M & Wang21 & 5.83 & \MgII & $47.621\pm0.004$ & $9.61\pm0.08$ & $-0.09\pm0.08$\\
				J0840+5624 & SDSS\_M & Wang21 & 5.82 & \MgII & $46.932\pm0.013$ & $9.37\pm0.12$ & $-0.54\pm0.12$\\
				J0842+1218 & SDSS\_M & Wang21 & 6.07 & \MgII & $47.196\pm0.005$ & $9.52\pm0.06$ & $-0.42\pm0.06$\\
				J0927+2001 & SDSS\_M & Shen19 & 5.77 & \CIV & $46.986\pm0.003$ & $9.73\pm0.10$ & $-0.86\pm0.10$\\
				J1030+0524 & SDSS\_M & Jiang07 & 6.31 & \MgII & $47.370\pm0.012$ & $9.46\pm0.05$ & $-0.19\pm0.05$\\
				J1044-0125 & SDSS\_M & Wang21 & 5.78 & \MgII & $47.311\pm0.006$ & $9.81\pm0.10$ & $-0.60\pm0.10$\\
				J1048+4637 & SDSS\_M & DeRosa11 & 6.20 & \MgII & $47.453\pm0.004$ & $9.49\pm0.14$ & $-0.14\pm0.14$\\
				J1137+3549 & SDSS\_M & Wang21 & 6.01 & \MgII & $47.282\pm0.008$ & $9.76\pm0.09$ & $-0.58\pm0.09$\\
				J1143+3808 & SDSS\_M & Wang21 & 5.80 & \CIV & $46.999\pm0.005$ & $9.73\pm0.08$ & $-0.83\pm0.08$\\
				J1148+5251 & SDSS\_M & Wang21 & 6.42 & \MgII & $47.533\pm0.004$ & $9.82\pm0.09$ & $-0.39\pm0.09$\\
				J1243+2529 & SDSS\_M & Wang21 & 5.84 & \CIV & $47.060\pm0.004$ & $9.84\pm0.05$ & $-0.88\pm0.05$\\
				J1250+3130 & SDSS\_M & Wang21 & 6.14 & \MgII & $46.988\pm0.005$ & $9.13\pm0.06$ & $-0.24\pm0.06$\\
				J1306+0356 & SDSS\_M & Jiang07 & 6.02 & \MgII & $47.400\pm0.011$ & $9.41\pm0.05$ & $-0.10\pm0.05$\\
				J1411+1217 & SDSS\_M & Jiang07 & 5.93 & \MgII & $47.200\pm0.011$ & $8.97\pm0.13$ & $0.13\pm0.13$\\
				J1602+4228 & SDSS\_M & Wang21 & 6.08 & \MgII & $47.210\pm0.009$ & $9.42\pm0.08$ & $-0.31\pm0.08$\\
				J1623+3112 & SDSS\_M & Wang21 & 6.25 & \MgII & $46.975\pm0.003$ & $9.32\pm0.15$ & $-0.45\pm0.15$\\
				J2310+1855 & SDSS\_M & Wang21 & 5.96 & \MgII & $47.464\pm0.005$ & $9.66\pm0.15$ & $-0.30\pm0.15$\\
				J0008-0626 & SDSS\_O & Wang21 & 5.93 & \MgII & $46.964\pm0.013$ & $9.19\pm0.07$ & $-0.33\pm0.07$\\
				J0028+0457 & SDSS\_O & Wang21 & 5.98 & \CIV & $46.969\pm0.020$ & $9.91\pm0.13$ & $-1.04\pm0.13$\\
				J0841+2905 & SDSS\_O & Wang21 & 5.95 & \MgII & $46.986\pm0.008$ & $9.40\pm0.19$ & $-0.51\pm0.19$\\
				J0850+3246 & SDSS\_O & Shen19 & 5.73 & \CIV & $47.195\pm0.003$ & $9.58\pm0.23$ & $-0.50\pm0.23$\\
				J1207+0630 & SDSS\_O & Wang21 & 6.03 & \MgII & $46.909\pm0.011$ & $9.53\pm0.08$ & $-0.72\pm0.08$\\
				J1257+6349 & SDSS\_O & Wang21 & 5.99 & \MgII & $46.739\pm0.013$ & $9.43\pm0.10$ & $-0.79\pm0.10$\\
				J1319+0950 & SDSS\_O & Schindler20 & 6.13 & \MgII & $47.249\pm0.002$ & $9.31\pm0.04$ & $-0.16\pm0.04$\\
				J1403+0902 & SDSS\_O & Shen19 & 5.79 & \CIV & $47.007\pm0.006$ & $9.17\pm0.37$ & $-0.28\pm0.37$\\
				J1630+4012 & SDSS\_O & Wang21 & 6.07 & \MgII & $46.760\pm0.007$ & $9.27\pm0.10$ & $-0.61\pm0.10$\\
				J0005-0006 & SDSS\_S82 & DeRosa11 & 5.85 & \MgII & $46.737\pm0.009$ & $8.03\pm0.06$ & $0.61\pm0.06$\\
				J0203+0012 & SDSS\_S82 & Shen19 & 5.71 & \CIV & $47.311\pm0.001$ & $10.05\pm0.12$ & $-0.85\pm0.12$\\
				J0303-0019 & SDSS\_S82 & DeRosa11 & 6.08 & \MgII & $46.579\pm0.007$ & $8.62\pm0.03$ & $-0.14\pm0.03$\\
				J0353+0104 & SDSS\_S82 & Wang21 & 6.06 & \MgII & $46.975\pm0.042$ & $9.32\pm0.17$ & $-0.45\pm0.17$\\
				J2054-0005 & SDSS\_S82 & Schindler20 & 6.04 & \MgII & $47.087\pm0.014$ & $9.02\pm0.12$ & $-0.03\pm0.12$\\
				\bottomrule
			\end{tabular}
			\begin{tablenotes}
				\small
				\item $^a$ The references of near-IR spectral fitting results: \citet{Jiang07BHM, DeRosa2011BHM, Wu15QSO, Schindler2020, Shen2019GNIRS, WangShu_etal_2021}
			\end{tablenotes}
		\end{threeparttable}
	\end{table*}
	
	\begin{table*}
		\begin{threeparttable}
			\caption{The $1/V_{\rm max}$(binned) BHMF}
			\label{tab:vmax-bhmf}
			\begin{tabular}{cccc}
				\toprule
				{Sample} & {$\mathrm{log}(M_{\mathrm{BH}})$} & {$\Delta\mathrm{log}(M_{\mathrm{BH}})$} & 
				{\makecell[ct]{$\mathrm{log}(\Phi(\lambda))$\\ $(\mathrm{Mpc^{-3} \mathrm{dex}^{-1}})$}} \\
				\midrule
				SDSS\_M&8.9&0.2&$-9.69^{+0.30}_{-\mathrm{inf}}$\\
				SDSS\_M&9.1&0.2&$-9.99^{+0.30}_{-\mathrm{inf}}$\\
				SDSS\_M&9.3&0.2&$-9.21^{+0.18}_{-0.32}$\\
				SDSS\_M&9.5&0.2&$-9.15^{+0.17}_{-0.28}$\\
				SDSS\_M&9.7&0.2&$-9.15^{+0.17}_{-0.28}$\\
				SDSS\_M&9.9&0.2&$-8.86^{+0.27}_{-0.84}$\\
				SDSS\_M&10.3&0.2&$-10.04^{+0.30}_{-\mathrm{inf}}$\\
				SDSS\_O&9.1&0.2&$-8.84^{+0.23}_{-0.53}$\\
				SDSS\_O&9.3&0.2&$-8.33^{+0.21}_{-0.42}$\\
				SDSS\_O&9.5&0.2&$-8.67^{+0.20}_{-0.37}$\\
				SDSS\_O&9.9&0.2&$-9.15^{+0.30}_{-\mathrm{inf}}$\\
				\bottomrule
			\end{tabular}
			\begin{tablenotes}
				\small
				\item Shown as orange and blue points in left panel of \autoref{fig:bhmf-1d} and \autoref{fig:lm-bhmf}.
			\end{tablenotes}
		\end{threeparttable}
	\end{table*}

	\begin{table*}
		\begin{threeparttable}
			\caption{The $1/V_{\rm max}$(binned) QLF}
			\label{tab:vmax-lf}
			\begin{tabular}{cccc}
				\toprule
				{Sample} & {$M_{1450}$} & {$\Delta M_{1450}$} & 
				{\makecell[ct]{$\mathrm{log}(\Phi(M_{1450}))$\\ $(\mathrm{Mpc^{-3} \mathrm{mag}^{-1}})$}} \\
				\midrule
				SDSS\_M&-26.25&0.5&$-9.61^{+0.18}_{-0.32}$\\
				SDSS\_M&-26.75&0.5&$-9.12^{+0.22}_{-0.46}$\\
				SDSS\_M&-27.50&1.0&$-9.65^{+0.13}_{-0.19}$\\
				SDSS\_M&-29.00&2.0&$-10.71^{+0.23}_{-0.54}$\\
				SDSS\_O&-25.75&0.5&$-9.24^{+0.23}_{-0.53}$\\
				SDSS\_O&-26.25&0.5&$-8.78^{+0.20}_{-0.36}$\\
				SDSS\_O&-26.75&0.5&$-9.25^{+0.23}_{-0.53}$\\
				SDSS\_O&-27.50&1.0&$-9.40^{+0.30}_{-15.85}$\\
				\bottomrule
			\end{tabular}
			\begin{tablenotes}
				\small
				\item Shown as orange and blue points in the right panel of \autoref{fig:bhmf-1d} and \autoref{fig:lm-lf}.
				\item These luminosities are calculated from $L_{\mathrm{bol}}$, see discussions in \autoref{sec:likelihood-method}.
			\end{tablenotes}
		\end{threeparttable}
	\end{table*}

	\begin{table*}
		\begin{threeparttable}
			\caption{The BHMF of model 3}
			\label{tab:bhmf-3}
			\begin{tabular}{ccccccc}
				\toprule
				\multicolumn{1}{c}{} &
				\multicolumn{3}{c}{Intrinsic distribution$^a$} &
				\multicolumn{3}{c}{Measured distribution$^b$} \\
				\cmidrule(l){2-4} \cmidrule(l){5-7}
				{log($M_{\mathrm{BH}}$)} & {$\log \Phi_-^d$} & {$\log \Phi^c$} & {$\log \Phi_+^d$} & {$\log \Phi_-^d$} & {$\log \Phi^c$} & {$\log \Phi_+^d$} \\
				 & & $(\mathrm{Mpc^{-3} \mathrm{dex}^{-1}})$ & & $(\mathrm{Mpc^{-3} \mathrm{dex}^{-1}})$ & & \\
				\midrule
				8.000 & -8.19 & -7.18 & -5.99 & -8.11 & -7.28 & -6.19 \\
				8.125 & -7.98 & -7.09 & -6.07 & -7.92 & -7.19 & -6.23 \\
				8.250 & -7.80 & -7.04 & -6.18 & -7.75 & -7.12 & -6.29 \\
				8.375 & -7.65 & -7.00 & -6.31 & -7.60 & -7.07 & -6.37 \\
				8.500 & -7.50 & -6.98 & -6.45 & -7.47 & -7.04 & -6.47 \\
				8.625 & -7.37 & -6.97 & -6.60 & -7.38 & -7.03 & -6.58 \\
				8.750 & -7.25 & -6.97 & -6.76 & -7.32 & -7.05 & -6.71 \\
				8.875 & -7.20 & -7.01 & -6.88 & -7.31 & -7.09 & -6.86 \\
				9.000 & -7.28 & -7.11 & -6.99 & -7.33 & -7.17 & -7.02 \\
				9.125 & -7.48 & -7.32 & -7.20 & -7.41 & -7.29 & -7.20 \\
				9.250 & -7.79 & -7.63 & -7.50 & -7.56 & -7.45 & -7.38 \\
				9.375 & -8.19 & -8.04 & -7.90 & -7.76 & -7.65 & -7.59 \\
				9.500 & -8.67 & -8.49 & -8.36 & -8.01 & -7.91 & -7.84 \\
				9.625 & -9.23 & -8.95 & -8.79 & -8.31 & -8.20 & -8.14 \\
				9.750 & -9.84 & -9.42 & -9.17 & -8.65 & -8.54 & -8.47 \\
				9.875 &-10.46 & -9.89 & -9.55 & -9.04 & -8.92 & -8.84 \\
				10.000 &-11.08 &-10.35 & -9.92 & -9.48 & -9.33 & -9.22 \\
				10.125 &-11.71 &-10.82 &-10.29 & -9.97 & -9.76 & -9.60 \\
				10.250 &-12.33 &-11.29 &-10.66 &-10.51 &-10.21 & -9.97 \\
				10.375 &-12.96 &-11.76 &-11.02 &-11.07 &-10.67 &-10.34 \\
				10.500 &-13.58 &-12.23 &-11.39 &-11.67 &-11.13 &-10.71 \\
				\bottomrule
			\end{tabular}
			\begin{tablenotes}
				\small
				\item $^a$ Plotted in \autoref{fig:lm-bhmf} as the black line with gray $1\sigma$ region.
				\item $^b$ Plotted in \autoref{fig:lm-bhmf} as the blue line with light blue $1\sigma$ region.
				\item $^c$ The best-fit model of model 3.
				\item $^d$ The lower and upper boundaries of the envelopes of the models with posterior probabilities larger than the 32th percentile in the fitting Monte Carlo Markov chain.
			\end{tablenotes}
		\end{threeparttable}
	\end{table*}

	\begin{table*}
		\begin{threeparttable}
			\caption{The QLF of model 3}
			\label{tab:lf-3}
			\begin{tabular}{ccccccc}
				\toprule
				\multicolumn{1}{c}{} &
				\multicolumn{3}{c}{Intrinsic distribution} &
				\multicolumn{3}{c}{Measured distribution} \\
				\cmidrule(l){2-4} \cmidrule(l){5-7}
				{log($M_{\mathrm{BH}}$)} & {$\log \Phi_-^d$} & {$\log \Phi^c$} & {$\log \Phi_+^d$} & {$\log \Phi_-^d$} & {$\log \Phi^c$} & {$\log \Phi_+^d$} \\
				& & $(\mathrm{Mpc^{-3} \mathrm{mag}^{-1}})$ & & $(\mathrm{Mpc^{-3} \mathrm{mag}^{-1}})$ & & \\
				\midrule
				-21.000 & -8.14 & -7.68 & -7.15 & -8.15 & -7.69 & -7.14 \\
				-21.500 & -8.07 & -7.69 & -7.27 & -8.07 & -7.69 & -7.26 \\
				-22.000 & -8.01 & -7.71 & -7.39 & -8.02 & -7.71 & -7.38 \\
				-22.500 & -7.98 & -7.74 & -7.52 & -7.98 & -7.75 & -7.51 \\
				-23.000 & -7.96 & -7.79 & -7.65 & -7.97 & -7.79 & -7.65 \\
				-23.500 & -8.00 & -7.86 & -7.77 & -8.00 & -7.86 & -7.76 \\
				-24.000 & -8.10 & -7.96 & -7.87 & -8.09 & -7.97 & -7.89 \\
				-24.500 & -8.26 & -8.11 & -8.03 & -8.25 & -8.12 & -8.04 \\
				-25.000 & -8.44 & -8.30 & -8.21 & -8.42 & -8.29 & -8.20 \\
				-25.500 & -8.68 & -8.54 & -8.45 & -8.63 & -8.50 & -8.41 \\
				-26.000 & -8.95 & -8.82 & -8.71 & -8.91 & -8.78 & -8.68 \\
				-26.500 & -9.27 & -9.13 & -9.03 & -9.22 & -9.09 & -9.00 \\
				-27.000 & -9.61 & -9.46 & -9.34 & -9.55 & -9.40 & -9.29 \\
				-27.500 &-10.03 & -9.81 & -9.67 & -9.91 & -9.73 & -9.60 \\
				-28.000 &-10.47 &-10.15 & -9.95 &-10.36 &-10.09 & -9.91 \\
				-28.500 &-10.93 &-10.50 &-10.24 &-10.82 &-10.44 &-10.19 \\
				-29.000 &-11.39 &-10.85 &-10.51 &-11.26 &-10.77 &-10.46 \\
				-29.500 &-11.86 &-11.21 &-10.80 &-11.72 &-11.13 &-10.75 \\
				-30.000 &-12.32 &-11.55 &-11.07 &-12.22 &-11.50 &-11.04 \\
				\bottomrule
			\end{tabular}
			\begin{tablenotes}
				\small
				\item The table headers have the similar meaning as those in \autoref{tab:bhmf-3}.
			\end{tablenotes}
		\end{threeparttable}
	\end{table*}

   \end{document}